\documentclass[aps,prl,twocolumn,showpacs,superscriptaddress]{revtex4-1}%
\usepackage{epsfig,dsfont,amssymb,amsmath,amsthm,amsfonts,amsbsy,mathrsfs}
\usepackage{graphicx}
\usepackage{amsmath}
\usepackage{amssymb}
\usepackage{multirow}
\usepackage[colorlinks]{hyperref}
\usepackage{color}
\usepackage{amsfonts}%
\providecommand{\U}[1]{\protect\rule{.1in}{.1in}}
\hypersetup{colorlinks,
linkcolor=blue,          citecolor=blue,        filecolor=blue,      urlcolor=blue           }
\newtheorem{theorem}{Theorem}

\newtheorem{definition}{Definition}

\newtheorem{lemma}{Lemma}

\newcommand{\bra}[1]{\langle#1|}
\newcommand{\ket}[1]{|#1\rangle}

\newcommand{\tr}{\mbox{tr}}
\newcommand{\poly}{\mbox{poly}}

\newcommand{\E}{\mathbb{E}}

\begin{document}

\preprint{APS/123-QED}
\title{Quantum Supremacy for Simulating A Translation-Invariant Ising Spin Model}

\author{Xun Gao}
\affiliation{Center for Quantum Information, Institute for Interdisciplinary Information Sciences, Tsinghua University, Beijing 100084, China} 
\author{Sheng-Tao Wang}
\affiliation{Department of Physics, University of Michigan, Ann Arbor, Michigan 48109, USA}
\author{L.-M. Duan}
\affiliation{Department of Physics, University of Michigan, Ann Arbor, Michigan 48109, USA}
\affiliation{Center for Quantum Information, Institute for Interdisciplinary Information Sciences, Tsinghua University, Beijing 100084, China}

\begin{abstract}
We introduce an intermediate quantum computing model built from
translation-invariant Ising-interacting spins. Despite being non-universal,
the model cannot be classically efficiently simulated unless the polynomial
hierarchy collapses. Equipped with the intrinsic single-instance-hardness property, a
single fixed unitary evolution in our model is sufficient to produce
classically intractable results, compared to several other models that rely on
implementation of an ensemble of different unitaries (instances). We propose a feasible experimental scheme to implement 
our Hamiltonian model using cold atoms trapped in a square optical lattice. We
formulate a procedure to certify the correct functioning of this quantum machine.
The certification requires only a polynomial number of local measurements
assuming measurement imperfections are sufficiently small.
\end{abstract}

\maketitle


A universal quantum computer is believed to be able to solve certain tasks
exponentially faster than the current computers \cite{nielsen2010quantum,
Ladd2010Quantum}. Over the past several decades, there has been tremendous
progress in both theoretical and experimental developments of a quantum
computer. In theory, pioneering quantum algorithms, including Shor's
factorization \cite{shor1994algorithms} and an algorithm for linear systems of
equations \cite{harrow2009quantum}, achieve exponential speedup compared with
the best-known classical algorithms. However, formidable experimental
challenges still lie ahead in building a universal quantum computer large
enough to demonstrate quantum supremacy. This calls for simpler tasks to
demonstrate exponential quantum speedup without the need for a universal machine.

Several intermediate computing models have been developed recently for this
purpose. Examples include boson sampling \cite{aaronson2011computational},
quantum circuits with commuting gates (IQP) \cite{bremner2011classical,
PhysRevLett.117.080501}, sparse and ``fault-tolerant" IQP \cite{Fujii2016Computational,bremner2016achieving}, the one-clean-qubit model \cite{morimae2014hardness,
fujii2014impossibility}, evolution of two-qubit commuting Hamiltonians
\cite{bouland2016complexity}, quantum approximate optimization algorithm \cite{2016arXiv160207674F} and random or universal quantum circuit \cite{boixo2016characterizing,fujii2016noise}. These models fall into the category of sampling
problems: the task of simulating the distribution sampled from the respective
quantum system is believed to be classically intractable. In particular, if a
classical computer can efficiently simulate the distribution to multiplicative
errors, the polynomial hierarchy, a generalization of $\mathsf{P}$ and
$\mathsf{NP}$ classes, will have to collapse to the third
level~\cite{arora2009computational, sm}, which is believed to be unlikely in
complexity theory. Several experiments  (e.g.~\cite{Broome2013Photonic, Spring2013Boson}) have been reported for realization of
boson sampling in small quantum systems using photons. However, the system size is still limited, which prohibits demonstration of quantum supremacy beyond classical tractability.

In this paper, we report three advancements towards demonstration of
exponential quantum speedup in intermediate computing models. First, we
formulate a new sampling model built from translation-invariant
Ising-interacting spins, with strong connection to simulation of natural
quantum many-body systems
\cite{feynman1982simulating,lloyd1996universal,Buluta2009Quantum,Cirac2012Goals}%
. Our model only requires nearest-neighbor Ising-type interactions. The state
preparation, the Hamiltonian and measurements are all constructed to be
translation-invariant. Similar to Refs.~\cite{aaronson2011computational,PhysRevLett.117.080501}, we prove the distribution
sampled from our model cannot be classically efficiently simulated based on
complexity theory results under reasonable conjectures \cite{toda1991pp,
han1997threshold,aaronson2005quantum,bremner2011classical}. 
An additional desirable feature of
our model, which we call the `single-instance-hardness' property, is that a single fixed circuit and measurement pattern are sufficient to produce a classically hard
distribution once the system size is fixed. This differs from typical sampling problems, where an
ensemble of instances (unitaries) with a large number of parameters is demanded for the hardness result to hold
\cite{aaronson2011computational,bremner2011classical,
PhysRevLett.117.080501,Fujii2016Computational,bremner2016achieving,morimae2014hardness,
fujii2014impossibility,bouland2016complexity,2016arXiv160207674F,boixo2016characterizing,fujii2016noise}. This feature offers a significant
simplification for experiments since proof of quantum supremacy for this model
requires implementation of only a single Hamiltonian and measurement pattern instead of a range of different realizations (typically an exponential number or even an infinite number). Ref.~\cite{aaronson2011computational} also discussed the single-instance-hardness possibility in an abstract quantum circuit language, but no explicit circuit has been given thus far. Second, we propose a feasible
experimental scheme to realize our model with cold atoms in optical lattices.
The state preparation, engineering of time evolution and measurement
techniques are achievable with the state-of-the-art technology. Unlike photonic systems, cold atomic systems are much easier to scale up and reach a system
size intractable to classical machines. Finally, we devise a scheme to certify
our proposed quantum machine based on extension of the techniques developed in Refs.~\cite{hangleiter2016direct, cramer2010efficient}. Certification of
functionality is critically important for a sampling quantum machine as a
correct sampling is hard to be verified. Our certification scheme only
requires a polynomial number of local measurements, assuming the measurement
imperfections are sufficiently small.

\begin{figure}[t]
\includegraphics[width=1\linewidth]{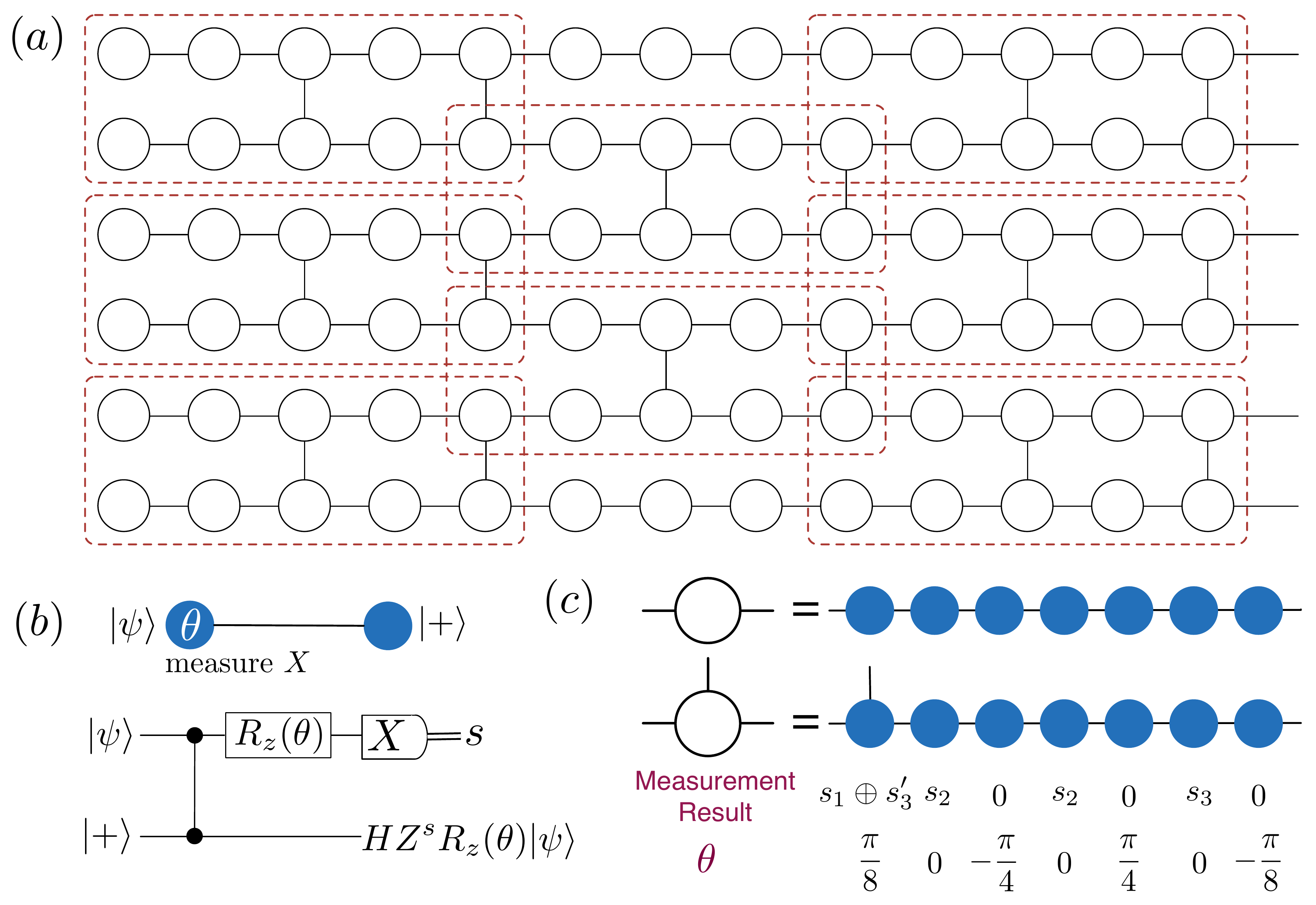} \caption{(a) The
brickwork state. Each circle represents a $|+\rangle$ state and each line
denotes a $CZ$ operation. (b) Propagation of the gate by measuring a qubit.
(c) Each white circle with varying rotation angles is replaced by seven
physical qubits with fixed rotation angles. The variation in the overall angle
is encoded into different measurement outcomes.}%
\label{fig:brickwork}%
\end{figure}

Before introducing our model, let us make more precise the two different error
requirements used in this paper. Suppose the distribution $\{q_{x}\}$ is
sampled from the quantum system with $q_{x}$ being the probability of
measuring the result $x$. Simulating $\{q_{x}\}$ to multiplicative errors
translates to finding another distribution $\{p_{x}\}$ such that
\begin{equation}
\forall x,|p_{x}-q_{x}|\leq\gamma q_{x}%
\end{equation}
with $\gamma<1/2$. This requirement seems too stringent for a classical
sampler \cite{bremner2011classical, aaronson2011computational}: even the
quantum device may not achieve such a physically unrealistic precision. A more sensible choice is the
variation distance error \cite{aaronson2011computational, PhysRevLett.117.080501,
fefferman2015power}
\begin{equation}
\sum_{x}|p_{x}-q_{x}|\leq\epsilon.\label{eq:tvd}%
\end{equation}
Other than physical motivation, another reason to use this quantification of
error lies in the equivalence between search and sampling problems under the
variation distance bound \cite{aaronson2014equivalence}: the separation
between classical and quantum samplers under this error requirement will
permit the quantum device to solve classically-intractable search problems
\cite{aaronson2011computational}. This will have broad practical applications
due to the ubiquity of search problems. For our Ising spin model, we will
prove that the distribution produced by the quantum sampler can be certified
by local measurements to variation distance errors, assuming the measurement
imperfections are sufficiently small.

Our model can be regarded as a special type of IQP with a constant circuit depth. A general IQP \cite{bremner2011classical, PhysRevLett.117.080501} consists of Ising interactions between any pairs of spins and with varying strengths, while the sparse IQP \cite{bremner2016achieving} has $O(\sqrt n\log n)$ depth. Note that we are able to achieve such a low depth while maintaining classical hardness with variation distance errors (Eq.~(\ref{eq:tvd})) because we use a different complexity conjecture of average-case hardness. Ref.~\cite{Fujii2016Computational} proposed another type of IQP in constant circuit depth on the Raussendorf-Harrington-Goyal (RHG) lattice~\cite{Raussendorf20062242}. In their model, the classical hardness result is guaranteed with multiplicative errors under some local noise below a threshold. Their Hamiltonian is also translation-invariant but the measurements are not. Thus, this model and the general IQP do not have the single-instance-hardness property. The general interactions in IQP and the three-dimensional structure of the RHG lattice may be difficult to realize in experiments.

\emph{Translation-invariant Ising model.---}Our main construction is based on measurement-based quantum computing models \cite{Briegel2009Measurement, Raussendorf2001A, Raussendorf2003Measurement}. We first introduce a
translation-invariant nonadaptive measurement-based quantum computation model with only one measurement basis required. With postselection, we show
that it can simulate universal quantum computation. Next, we reinterpret the
measurement-based model as a sampling model based on quantum simulation of
two-dimensional (2D) spins with translation-invariant Ising interactions and
local magnetic fields. It has been known that if a sampling model with
postselection can simulate universal quantum computation, it will be hard to
simulate classically with multiplicative error bounds unless the polynomial
hierarchy collapses to the third level \cite{bremner2011classical,
morimae2014hardness, bouland2016complexity}. We therefore conclude that our
quantum Ising model will be classically intractable if the polynomial
hierarchy does not collapse \cite{sm}.

Consider the brickwork state shown in Fig.~\ref{fig:brickwork}(a), which has
been used for universal blind quantum computation
\cite{broadbent2009universal}. Each circle represents a qubit prepared in the
state $|+\rangle=(|0\rangle+|1\rangle)/\sqrt{2}$. A line connecting two
neighboring circles denotes a controlled-Z operation on the qubits. As
illustrated in Fig.~\ref{fig:brickwork}(b), a measurement on one qubit in $X$
basis with measurement result $s$ implements a gate $HZ^{s}R_{z}(\theta)$,
where $H$ is the Hadamard gate and $R_{z}(\theta)=e^{-i\theta Z/2}$ denotes a
rotation on a single qubit. Ref.~\cite{broadbent2009universal} proved that the
model supports universal quantum computation given proper rotation angles
$\theta$ and measurement results $s$ (see Supplemental Material \cite{sm} for
details). An important attribute of this model is that the graph structure and
measurement patterns are independent of the computation. We further improve
the model by making the angles $\theta$ translation-invariant. In terms of the
sampling problem, this modification gives rise to the advantage of the single-instance-hardness property. It differs from other existing sampling problems, such
as boson sampling, wherein an average over random quantum circuits is needed
for the classical hardness result to hold.

To fix the angle pattern, we use seven qubits to replace one white circle
(Fig.~\ref{fig:brickwork}(c)). The primary goal is to encode rotation angle
values into measurement outcomes, so that measurement postselection
effectively realizes all necessary rotation angles. The basic building block
is
\begin{equation}
HZ^{s}HR_{z}\left(  -\frac{\theta}{2}\right)  HZ^{s}HR_{z}\left(  \frac
{\theta}{2}\right)  =R_{z}^{s}(\theta)
\end{equation}
which can be realized by measuring four connecting qubits in $X$ basis with
rotation angles $\theta/2,0,-\theta/2,0$ and postselecting the results to be
$0,s,0,s$. This equality furnishes a mechanism to conditionally perform the
rotation $R_{z}(\theta)$ based on the measurement result $s$. Because of the
Solovey-Kitaev theorem \cite{kitaev1997quantum}, it is sufficient to implement
$HR_{z}(k\pi/4),k\in\{0,\cdots,7\}$ for universal computation \cite{sm}.
Writing $k=s_{1}s_{2}s_{3},s_{i}\in\{0,1\}$ in binary form, we have
\begin{align*}
Z^{s_{3}}HR_{z}\left(  \frac{k\pi}{4}\right)  Z^{s_{3}^{\prime}}\!= &
Z^{s_{3}}HR_{z}^{s_{1}}(\pi)R_{z}^{s_{2}}\left(  \frac{\pi}{2}\right)
R_{z}^{s_{3}}\left(  \frac{\pi}{4}\right)  Z^{s_{3}^{\prime}}\\
= &  HR_{z}\left(  -\frac{\pi}{8}\right)  HZ^{s_{3}}HR_{z}\left(  \frac{\pi
}{4}\right)  HZ^{s_{2}}\\
&  HR_{z}\left(  -\frac{\pi}{4}\right)  HZ^{s_{2}}HZ^{s_{1}+s_{3}^{\prime}%
}R_{z}\left(  \frac{\pi}{8}\right)  .
\end{align*}
The extra term $Z^{s_{3}}$ can be absorbed into the following gate and
$Z^{s_{3}^{\prime}}$ is left from the previous gate. Postselecting the
measurement results as $s_{1}\oplus s_{3}^{\prime},s_{2},0,s_{2},0,s_{3},0$
with rotation angles $\pi/8,0,-\pi/4,0,\pi/4,0,-\pi/8$, we can implement the
gates $HR_{z}(k\pi/4)$ with $k=s_{1}s_{2}s_{3}$. 

We now recast the nonadaptive measurement-based computation model as a
sampling problem. A distribution can be sampled by measuring each spin in Fig.~\ref{fig:brickwork} in $X$ basis. The above procedure is only used to prove the universality
of the nonadaptive measurement-based model with a fixed circuit under
postselection. We remark that neither postselection nor adaptive measurements
are required for sampling the distribution. The circuit can be implemented by
a unitary time evolution under a local Hamiltonian
\begin{equation}
\mathcal{H}=-\sum_{\left\langle i,j\right\rangle }JZ_{i}Z_{j}+\sum_{i}%
B_{i}Z_{i}\label{eq:Ham}%
\end{equation}
starting from the initial state $|+\rangle^{\otimes m\times n}$, with $m\times
n$ being the number of spins. The second term imprints local rotation angles
since $e^{-iB_{i}Z_{i}}=R_{z}(\theta_{i})$, where $B_{i}=\theta_{i}/2$
characterizes the local Zeeman field strength on spin $i$. The evolution time
and the reduced Planck constant $\hbar$ are set to unity. The first term
performs the controlled-Z operations with $J=\pi/4$, where $\left\langle
i,j\right\rangle $ represents nearest-neighbor pairs connected by a line in
Fig.~\ref{fig:brickwork}. This can be seen as
\begin{align}
CZ_{ij} &  =e^{i\pi|1\rangle\langle1|_{i}\otimes|1\rangle\langle1|_{j}%
}=e^{i\pi/4(I_{i}-Z_{i})\otimes(I_{j}-Z_{j})}\nonumber\\
&  =e^{i\pi/4}e^{-i\pi/4I_{i}\otimes Z_{j}}e^{-i\pi/4Z_{i}\otimes I_{j}%
}e^{i\pi/4Z_{i}\otimes Z_{j}}.
\end{align}
The two local magnetic field terms in the equation above can be absorbed into
rotation angles, without changing Fig.~\ref{fig:brickwork}(c) (see
Supplemental Material \cite{sm}). The distribution sampled from this fixed 2D
Ising model cannot be simulated by a classical computer in polynomial time to
multiplicative errors unless the polynomial hierarchy collapses.

\emph{Implementation proposal with cold atoms.---}The Hamiltonian in
Eq.~\eqref{eq:Ham} exhibits a few properties that make it amenable for
experimental implementation. First of all, it only consists of commuting
terms, so in experiment one can choose to break up the Hamiltonian and apply
simpler terms in sequence. Second, the state preparation, the Hamiltonian and
measurements are all translation-invariant. This may greatly simplify the
implementation for setups that can engineer the required unit cell. Another
merit of our model originates from the single-instance-hardness feature. It ensures the
sampling distribution after a single fixed unitary operation is already hard
to simulate classically.

Here, we put forward a feasible experimental scheme based on cold atoms in
optical lattices. A major difficulty arises from the special geometry required
in the brickwork state. We propose to circumvent this problem by starting from
the 2D cluster state (square lattice geometry) and reducing it to the
brickwork state. In theory, this can be achieved by the \textquotedblleft
break" and \textquotedblleft bridge" operations with measurement postselection
as shown in Fig.~\ref{fig:break_bridge} (see Supplemental Material \cite{sm}
for more details). In experiment, postselection is again unnecessary with
regard to sampling, but one incurs an additional cost of measuring in both $X$
and $Z$ basis (the measurement pattern is still translation-invariant though).
As a by-product, this procedure offers a concrete single-instance-hardness protocol to
produce classically non-simulatable distribution from the cluster state.

\begin{figure}[t]
\includegraphics[width=0.85\linewidth]{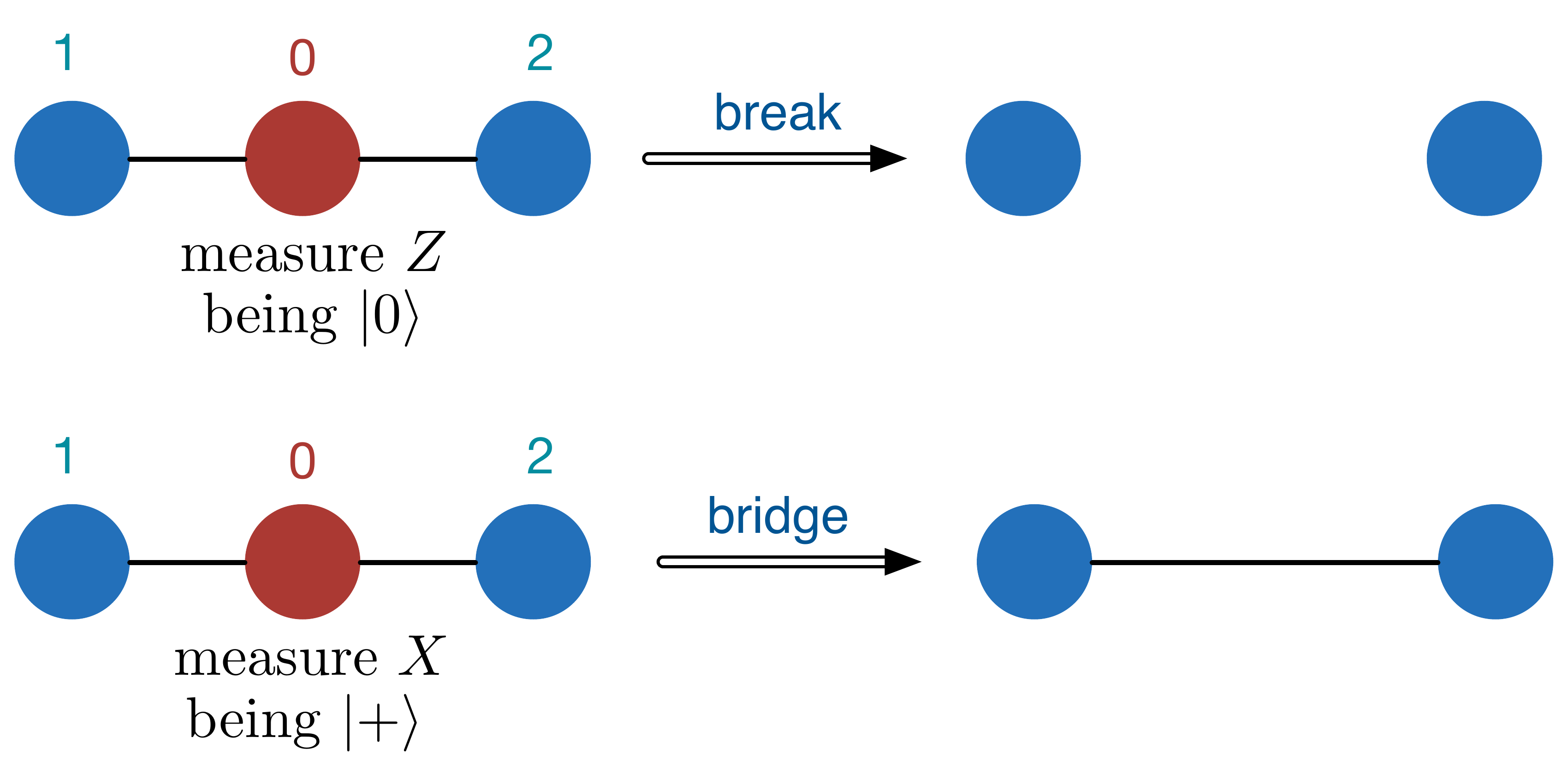} \caption{Break and
bridge operations. Qubit 0 is first rotated by $R_{z}(\pi/2)$ before measured
in $Z$ and $X$ basis respectively to perform the break and bridge operations.}%
\label{fig:break_bridge}%
\end{figure}

A complete experimental procedure is as follows. First, create a
Mott-insulator state of cold atoms in 2D optical lattices with a central core
of unit filling. One atom with two relevant atomic levels (e.g.,
$|F=1,m_{F}=-1\rangle$ and $|F=2,m_{F}=-2\rangle$ hyperfine levels of $^{87}%
$Rb atoms) can be trapped in each site forming a square lattice of qubits. A
2D cluster state can be created in a single operational step by controlled
collisional interaction \cite{Jaksch1999Entanglement, Mandel2003Controlled}.
The basic idea involves entangling neighboring atoms by spin-dependent
transport together with controlled on-site collisions, which has been realized
in experiment \cite{Mandel2003Controlled}. After generating the cluster state,
one needs to impose the rotation angle pattern onto each qubit. This requires
the ability to address individual atoms with diffraction-limited performance.
Single-site addressing is currently one of the state-of-the-art quantum
control techniques in cold atom experiments \cite{Weitenberg2011Nature,
Zupancic2016Ultra}. In particular, by using a digital micro-mirror device, it
is possible to engineer holographic beam shaping with arbitrary amplitude and
phase control \cite{Zupancic2016Ultra}. To imprint the individual phases, one
can make use of spin-dependent AC Stark shifts \cite{Weitenberg2011Nature}
with beam amplitude patterns given by the rotation angles. The amplitude
hologram controls the strength $B_{i}$ and realizes the second term in the
Hamiltonian in Eq.~\eqref{eq:Ham}. Finally, spin measurements can be performed
on each site, with single-site-resolved imaging techniques
\cite{Bakr2010Probing, Sherson2010Single}. Because some spins have to be
measured in $Z$ basis, they should be rotated by individual addressing
techniques before all atoms can be measured in $X$ basis.

\emph{Simulation and certification with variation distance errors.---}So far,
we have shown that our Ising spin model is classically intractable with
multiplicative error bounds. Similar to what have been attained in boson
sampling \cite{aaronson2011computational} and IQP \cite{PhysRevLett.117.080501},
we can also prove classical hardness to variation distance error bounds if we
assume the \textquotedblleft worst-case\textquotedblright\ hardness result can
be extended to \textquotedblleft average-case\textquotedblright. More
specifically, let us define the partition function of
\begin{equation}
\mathcal{H}_{x}=\mathcal{H}+\frac{\pi}{2}\sum_{i}x_{i}Z_{i},\text{ where
}x_{i}\in\{0,1\}
\end{equation}
to be $\mathcal{Z}_{x}=\mbox{tr}\left(  e^{-\beta\mathcal{H}_{x}}\right)  $,
setting the imaginary temperature unit as $\beta\equiv1/k_{B}T=i$. In Supplemental
Material \cite{sm}, we prove that approximating $|\mathcal{Z}_{x}|^{2}/2^{mn}$
by $\widetilde{|\mathcal{Z}_{x}|^{2}}/2^{mn}$ to a mixture of multiplicative
and additive errors such that
\begin{equation}
\left\vert \frac{\widetilde{|\mathcal{Z}_{x}|^{2}}}{2^{mn}}-\frac
{|\mathcal{Z}_{x}|^{2}}{2^{mn}}\right\vert \leq\frac{1}{\mbox{poly}(n)}%
\frac{|\mathcal{Z}_{x}|^{2}}{2^{mn}}+\frac{\epsilon}{\delta}%
(1+o(1))\label{eq:mix_error}%
\end{equation}
with $\epsilon/\delta<1/2$ is \#\textsf{P}-hard in the worst-case. Our
classical intractability result requires lifting the \#\textsf{P}-hardness of
the estimation from the worst-case to the average-case: picking any $1-\delta$
fraction of instances $x$, it is still \#\textsf{P}-hard. This conjecture is
similar to the one used in Ref.~\cite{PhysRevLett.117.080501} except that they
reduced the mixture of errors to simply multiplicative errors. 
All the known classically intractable quantum sampling models with variation distance errors require a similar average-case complexity conjecture.

Thus, with reasonable assumptions, our Ising spin model is also classically
intractable with variation distance bounds. Using techniques similar to those
in Refs.~\cite{hangleiter2016direct, cramer2010efficient}, we can in addition
certify the correct functioning of a quantum device, with only a polynomial
number of local measurements. Suppose $\{q_{x}^{\prime}\}$ is the distribution
sampled from our quantum device with the final state $\rho^{\prime}$ (state
before measurement); the ideal ones are denoted as $\{q_{x}\}$ and $\rho$. The
total variation distance between distributions $\{q_{x}\}$ and $\{q_{x}%
^{\prime}\}$ can be bounded by~\cite{nielsen2010quantum}:
\begin{equation}
\sum_{x}|q_{x}-q_{x}^{\prime}|\leq D(\rho,\rho^{\prime}),
\end{equation}
where $D(\rho,\rho^{\prime})=\mbox{tr}(|\rho-\rho^{\prime}|)/2$ is the trace
distance between states $\rho$ and $\rho^{\prime}$. Hence, if we can bound the
trace distance $D(\rho,\rho^{\prime})<\epsilon$, we can also bound the total
variation distance. Note, however, this does not allow us to estimate $q_{x}$
in experiment: statistical errors always kick in to thwart any polynomial-time
efforts to estimate the distribution due to the exponential suppression of
some $q_{x}$. We bypass statistical errors by assuming the correctness of
quantum mechanics. To sample from $\{q_{x}^{\prime}\}$ in experiment though,
measurement imperfections may cause deviations in variation distance. However,
if measurement imperfections on each spin are local and bounded by
$O(\epsilon/(mn))$ \cite{sm}, we can still correctly certify the quantum
device. Below, we show how to bound $D(\rho,\rho^{\prime})$ by a polynomial
number of local measurements.

As a graph state, the brickwork state in Fig.~\ref{fig:brickwork}(a)(c), is
the unique ground state of the 4-local Hamiltonian
\begin{equation}
H_{\text{brickwork}}=\sum_{i}\frac{I-X_{i}\prod_{j\in\text{neighbor of }%
i}Z_{j}}{2}.
\end{equation}
Each qubit $i$ is connected to at most three neighboring ones, and the energy
gap from the ground state is $1$. The ideal state $\rho$ is the brickwork
state acted by some single qubit rotations $R_{z}(\theta_{i})$. It is
therefore the unique ground state of the Hamiltonian
\begin{align*}
H_{\text{brickwork}}^{\prime} &  =\prod_{i}R_{z}(\theta_{i}%
)H_{\text{brickwork}}\prod_{j}R_{z}^{\dag}(\theta_{j})\\
&  =\sum_{i}\frac{I-R_{z}(\theta_{i})X_{i}R_{z}^{\dag}(\theta_{i})\prod
_{j\in\text{neighbor of }i}Z_{j}}{2}.
\end{align*}
This Hamiltonian is still $4$-local, with ground state energy gap $1$. Using
the weak-membership quantum state certification protocol in
Ref.~\cite{hangleiter2016direct}, one can measure each local term of
$H_{\text{brickwork}}^{\prime}$ by a polynomial number of times to obtain a
good estimation of $\langle H_{\text{brickwork}}^{\prime}\rangle$ averaged
over $\rho^{\prime}$. The estimation will be efficient due to Hoeffding's
bound and the finite norm of each local term. Since the ground state energy
gap is constant, $\langle H_{\text{brickwork}}^{\prime}\rangle>0$ implies a
finite component of excited states is present in $\rho^{\prime}$. Conversely,
a small $\langle H_{\text{brickwork}}^{\prime}\rangle$ will be able to bound
$D(\rho,\rho^{\prime})$. More quantitatively, we show in Supplemental Material
\cite{sm} that with confidence level $1-2^{-O(r)}$, using $O(m^{2}%
n^{2}r/\epsilon^{4})$ measurements on each local term is sufficient to certify
$\sum_{x}|q_{x}-q_{x}^{\prime}|\leq\epsilon$, provided the measurement
imperfections on each spin are bounded by $O(\epsilon/(mn))$. Similar hardness
and certification results hold if we start from the cluster state as in our
experimental proposal~\cite{sm}. In that case, 5-local measurements are needed.

The IQP certification protocol developed in Ref.~\cite{hangleiter2016direct}
requires a much stronger quantum simulator than the IQP simulator itself since
they need to generate all the history states \cite{kitaev2002classical}. In
contrast, our certification protocol only requires preparing the state
$\rho^{\prime}$ itself. This is relevant in light of demonstrating quantum
supremacy \cite{preskill2012quantum} using practical quantum many-body
systems, instead of resorting to a universal quantum simulation device.

\emph{Discussion.---}In summary, we have introduced a translation-invariant Ising spin model and shown that it is classically intractable unless the
polynomial hierarchy collapses. Because our average-case conjecture bypasses the anticoncentration property used in Refs.~\cite{aaronson2011computational,PhysRevLett.117.080501,bremner2016achieving}, the classical simulability result under constant-strength local noise \cite{bremner2016achieving} may not apply to our model. Whether our model is robust to noise requires further analysis. There is also a natural connection between our model and sampling models of random quantum circuits such as the one in Ref.~\cite{boixo2016characterizing}: measurement on qubits in the first $n-1$ columns in our model corresponds to choosing one instance of a random circuit due to the relation between our model and measurement-based quantum computing. With the advantageous single-instance-hardness property, the amenability to experimental implementation and certification of the quantum machine, we develop a full picture of using our model to
demonstrate quantum supremacy. This may shed light on the likely exponential
gap in computational power between a classical and a quantum machine.

\begin{acknowledgments}
X. G. is grateful to Man-Hong Yung and Mingji Xia for helpful discussions that inspire this work. S. T. W. thanks Ruichao Ma for useful discussions related to the experimental proposal. We also thank Michael J. Bremner and Ashley Montanaro for insightful discussions and suggestions. This work was supported by the Ministry of Education of China and Tsinghua University. L. M. D. and S. T. W. acknowledges in addition support from the AFOSR MURI program and the IARPA program.
\end{acknowledgments}

\begin{widetext}

\section{Supplemental Material}

In this Supplemental Material, we provide more details on the proof that our model is classically intractable to multiplicative errors based on some complexity results. We also show that the hardness result can be extended to variation distance error bounds if the worse-case results can be strengthened to the average-case. In addition, we demonstrate how to certify the quantum sampler if measurement imperfections can be made sufficiently small.

\subsection{Related Complexity Results}
In this section, we review some definitions and results on complexity theory related to our discussions in the main text of our paper. 
We adopt the same definitions as in Ref.~\cite{arora2009computational}, which includes more detailed discussions on these complexity classes. The concept of language $L$ (a subset of the string $\{0,1\}^*$) is used to formalize decision problems (of which solution can only be true or false). We call an instance of the problem as $x$; if the solution of $x$ is true, $x\in L$, otherwise $x\notin L$.

Before introducing those complexity classes directly used in this paper, we give a formal definition of the well known class \textsf{NP}. Intuitively, \textsf{NP} is the set of problems for which the ``yes'' solutions can be efficiently verified by a classical computer.
\begin{definition}[\textsf{NP}: nondeterministic polynomial]
A language $L$ is in \textsf{\emph{NP}} if there exists a polynomial $p$ and a polynomial time classical Turing Machine $M$ such that for every $x\in\{0,1\}^*$
$$x\in L\Leftrightarrow\exists u\in\{0,1\}^{p(|x|)} \mbox{ s.t. }M(x,u)=1.$$
\end{definition}

Polynomial hierarchy is in some sense a generalization of \textsf{NP}.
\begin{definition}[$\mathsf{\Sigma}_i^p$,\textsf{PH}: polynomial hierarchy] For $i\ge1$, a language $L$ is in $\mathsf{\Sigma}_i^p$ if there exists a polynomial $q$ and a polynomial time classical Turing Machine $M$ such that for every $x\in\{0,1\}^*$
\begin{eqnarray*}
x\in L&\Leftrightarrow&\exists u_1\in\{0,1\}^{q(|x|)}\forall u_2\in\{0,1\}^{q(|x|)}\cdots Q_i u_i\\
&&\in\{0,1\}^{q(|x|)}
\mbox{ s.t. }M(x,u_1,u_2,\cdots,u_i)=1,
\end{eqnarray*}
where $Q_i$ denotes $\forall$ or $\exists$ depending on whether $i$ is even or odd, respectively. And
$$\mathsf{PH}=\bigcup_i\mathsf{\Sigma}_i^p.$$
\end{definition}
Note that $\mathsf{NP}=\mathsf{\Sigma}_1^p$ and one can generalize $i$ to $0$ such that $\mathsf{P}=\mathsf{\Sigma}_0^p$. Clearly, $\mathsf{\Sigma}_i^p\subseteq\mathsf{\Sigma}_{i+1}^p\subseteq\mathsf{PH}$. Most computer scientists believe $\mathsf{P}\ne\mathsf{NP}$. A generalization of this conjecture is that for every $i$, $\mathsf{\Sigma}_i^p$ is strictly contained in $\mathsf{\Sigma}_{i+1}^p$,  which means $\mathsf{\Sigma}_{i}^p\ne\mathsf{\Sigma}_{i+1}^p$. It can also be stated as ``the polynomial hierarchy does not collapse". This conjecture is often used in complexity theory.

There is another way to generalize the class $\mathsf{NP}$. According to the above definition, it only requires knowing whether there exists at least one witness such that the Turing machine accepts. Counting problems need to compute the number of witnesses. This class is defined as
\begin{definition}[\textsf{\#P}] 
A function $f$ is in \textsf{\emph{\#P}} if there exists a polynomial $q$ and a polynomial time classical Turing machine $M$ such that for every $x\in\{0,1\}^*$
$$f(x)=\#\{y\in\{0,1\}^{q(|x|)}:M(x,y)=1\}.$$
\end{definition}

The following two complexity classes are directly related to sampling problems. One complexity class is \textsf{postBQP} defined in Ref.~\cite{aaronson2005quantum}. This complexity class characterizes the computational power of a universal quantum computer given the ability to do postselection. The other is a classical analog, \textsf{postBPP}, defined in Ref.~\cite{han1997threshold}.
\begin{definition}[\textsf{postBQP},\textsf{postBPP}]\label{def:post}
A language $L$ is in $\mathsf{postBQP}/\mathsf{postBPP}$ if there exists a uniform (which means can be generated by a classical polynomial Turing Machine) family of polynomial size quantum/classical circuits $Q_{n}/C_n$ such that for every $x\in\{0,1\}^*$, after applying $Q_{n}/C_n$ to the state
\begin{itemize}
\item the probability measuring registers $P/\widetilde P$ (called postselection registers) in the state $\ket{0\cdots0}/0\cdots0$ is nonzero;
\item if $x\in L$, then conditioned on measuring $P/\widetilde P$ on state $\ket{0\cdots 0}/0\cdots0$, the probability measuring the output register on state $\ket 1/1$ is at least $a$ (completeness error);
\item if $x\notin L$, then conditioned on measuring $P/\widetilde P$ on state $\ket{0\cdots 0}/0\cdots0$, the probability measuring the output register on state $\ket 1/1$ is at most $b$ (soundness error).
\end{itemize}
where $a-b>1/\mbox{\emph{poly}}(n)$.
\end{definition}

Some relations between these classes are included in the following theorem.  
\begin{theorem}\label{thm:classchain}
The first is Toda's theorem \cite{arora2009computational}, the second is proved in Ref.~\cite{aaronson2005quantum}, and the third is proved in Ref.~\cite{han1997threshold}:
\begin{eqnarray*}
\mathsf{PH}&\subseteq&\mathsf{P}^\mathsf{\#P}\\
\mathsf{P}^\mathsf{\#P}&=&\mathsf{P}^\mathsf{postBQP}\\
\mathsf{postBPP}&\subseteq&\mathsf{\Sigma}_{3}^p.
\end{eqnarray*}
\end{theorem}

In order to simulate \textsf{postBQP} by postselection, we need to define an output register $O/\widetilde O$ which gives the result of the decision problem, and a postselection register $P/\widetilde P$ of which the result is postselected to be some string of $\{0,1\}$. The key point is that we can change the definition slightly without changing the classes \textsf{postBQP} and \textsf{postBPP}: replacing the result of the register $P/\widetilde P$ by
\begin{equation}
\ket{0\cdots0}/0\cdots0\longrightarrow \ket{s_1\cdots s_{m\times n-1}}/s_1\cdots s_{m\times n-1}.
\end{equation}
This is crucial to our result.

Suppose the result in the output register is $x$. Classical simulability with multiplicative error implies
\begin{equation}\label{eq:mpe}
\frac{1}{c}q_{xs_1\cdots s_{m\times n-1}}\le p_{xs_1\cdots s_{m\times n-1}}\le cq_{x\&s_1\cdots s_{m\times n-1}}
\end{equation}
where the probability $\{q\}$ is sampled by our model, denoted as \textsf{Ising} and $\{p\}$ is sampled by a classical polynomial probabilistic Turing machine, shorted as \textsf{BPP};
the first digit $x$ is in the register $O/\widetilde O$ and other digits $s_1\cdots s_{m\times n-1}$ are in the register $P/\widetilde P$; This is equivalent to the definition of Eq.~(2) in the main text if we choose $\gamma=\min(1-1/c,c-1)$.

With postselection, we can define \textsf{postIsing}. The output probability is
$$
R(x)\equiv\frac{q_{xs_1\cdots s_{m\times n-1}}}{q_{0s_1\cdots s_{m\times n-1}}+q_{1s_1\cdots s_{m\times n-1}}}.
$$
The output probability of the corresponding \textsf{postBPP} is
$$
\widetilde R(x)\equiv\frac{p_{xs_1\cdots s_{m\times n-1}}}{ p_{0\&s_1\cdots s_{m\times n-1}}+p_{1\&s_1\cdots s_{m\times n-1}}}.
$$
According to the definition of multiplicative error Eq.~(\ref{eq:mpe}), we have
$$
\frac{1}{c^2}R(x)\le\widetilde R(x)\le c^2R(x),
$$
With this inequality and if $c<\sqrt{2}$ (so $\gamma<1/2$),
$$
|\widetilde R(0)-\widetilde R(1)|>0 \, \mbox{(not scaling with the problem size)}\Rightarrow |R(0)-R(1)|>0.
$$
This condition means that if there is a gap between completeness and soundness error in \textsf{Ising}, there will also be a gap for the \textsf{BPP} simulator:
\begin{equation}
\mathsf{postIsing}\subseteq\mathsf{postBPP}.
\end{equation}
If we can further prove
\begin{equation}
\mathsf{postBQP}\subseteq\mathsf{postIsing}.
\end{equation}
which means \textsf{Ising} with postselection can simulate universal quantum computer.
Combined with theorem \ref{thm:classchain}, we have
\begin{equation}
\mathsf{PH}\subseteq\mathsf{P}^\mathsf{\#P}=\mathsf{P}^\mathsf{postBQP}=\mathsf{P}^\mathsf{postIsing}\subseteq
\mathsf{P}^\mathsf{postBPP}\subseteq\mathsf{\Sigma}_{3}^p,
\end{equation}
which means the polynomial hierarchy collapses to the third level.
This contradicts with the generalization of the \textsf{P}$\ne$\textsf{NP} conjecture
\begin{equation}
\mathsf{\Sigma}_{3}^p\varsubsetneq \mathsf{PH}.
\end{equation}
Here, we adopt the same idea of proof as in Ref.~\cite{bremner2011classical}.

\subsection{Universal Quantum Computation with the Brickwork State}

Ref.~\cite{broadbent2009universal} has given the proof of universality. For completeness, we briefly review the result. Fig.~\ref{figSup:ucircuit}(a) shows how to choose different angles to get any single qubit gates and the CNOT gate. They are known to be universal. Fig.~\ref{figSup:ucircuit}(b) shows how to combine two qubit gates together to implement universal quantum computation.

\begin{figure}[t]
    \includegraphics[width=1\linewidth]{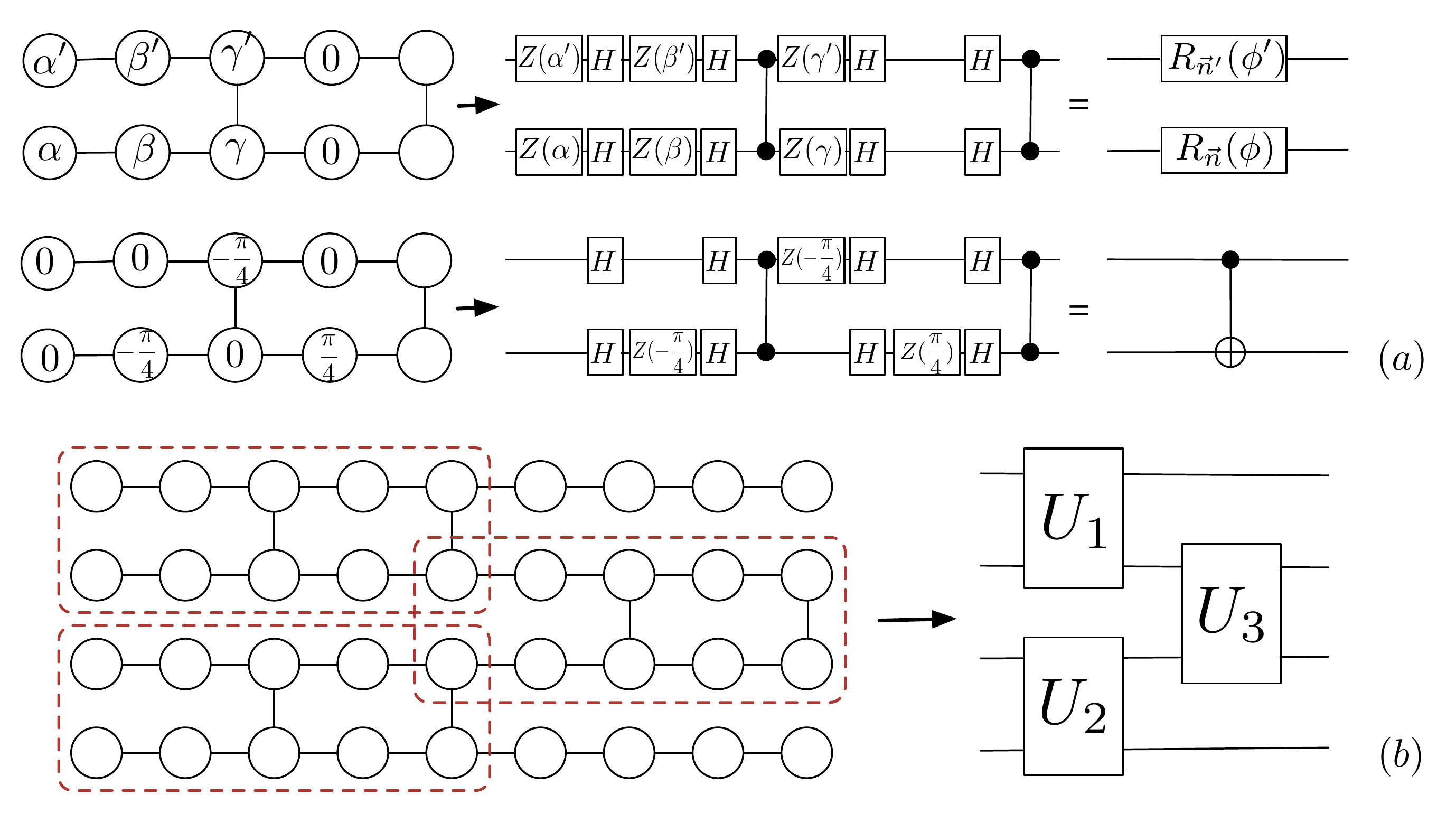}\caption{Implementing universal quantum computation with the brickwork state. These figures are similar to the ones in Ref.~\cite{broadbent2009universal}.}
    \label{figSup:ucircuit}
\end{figure}

\section{Magnetic Field in the Ising Spin Model}

In this section, we show that the extra local magnetic fields can be absorbed into the magnetic fields of Fig.~1(c) of the main text of our paper. We have three separate cases:
\begin{itemize}
\item For those spins that only couple with one other spin, there is an extra magnetic field $R_{z}(\pi/2)$. This spin must be on the left or the right boundary of the brickwork state. We can regard it as an ordinary unitary operation acting on the input. It can be eliminated by acting $R_{z}(-\pi/2)$ on the remaining quantum circuits. 
\item For those spins that couple with two other spins, there is an extra magnetic field $R_{z}(\pi)$. These spins will be acted on by an extra $Z$ gate. It can be eliminated by flipping the measurement result.
\item For those spins that couple with three other spins, there is an extra magnetic field $R_{z}(3\pi/2)$. These spins must have a vertical coupling; according to Fig.~1(c) of the main text, we can make the rotation angle $\theta$ on those spins to be $\pi/8+3\pi/2=\pi/8-\pi/2 \mod 2\pi$. It can be eliminated by flipping the measurement result from $s_2$ to  $s_2\oplus1$ and from $s_3$ to $s_3\oplus s_2$.
\end{itemize}

\subsection{break and bridge operations}

\begin{figure}[t]
    \includegraphics[width=1\linewidth]{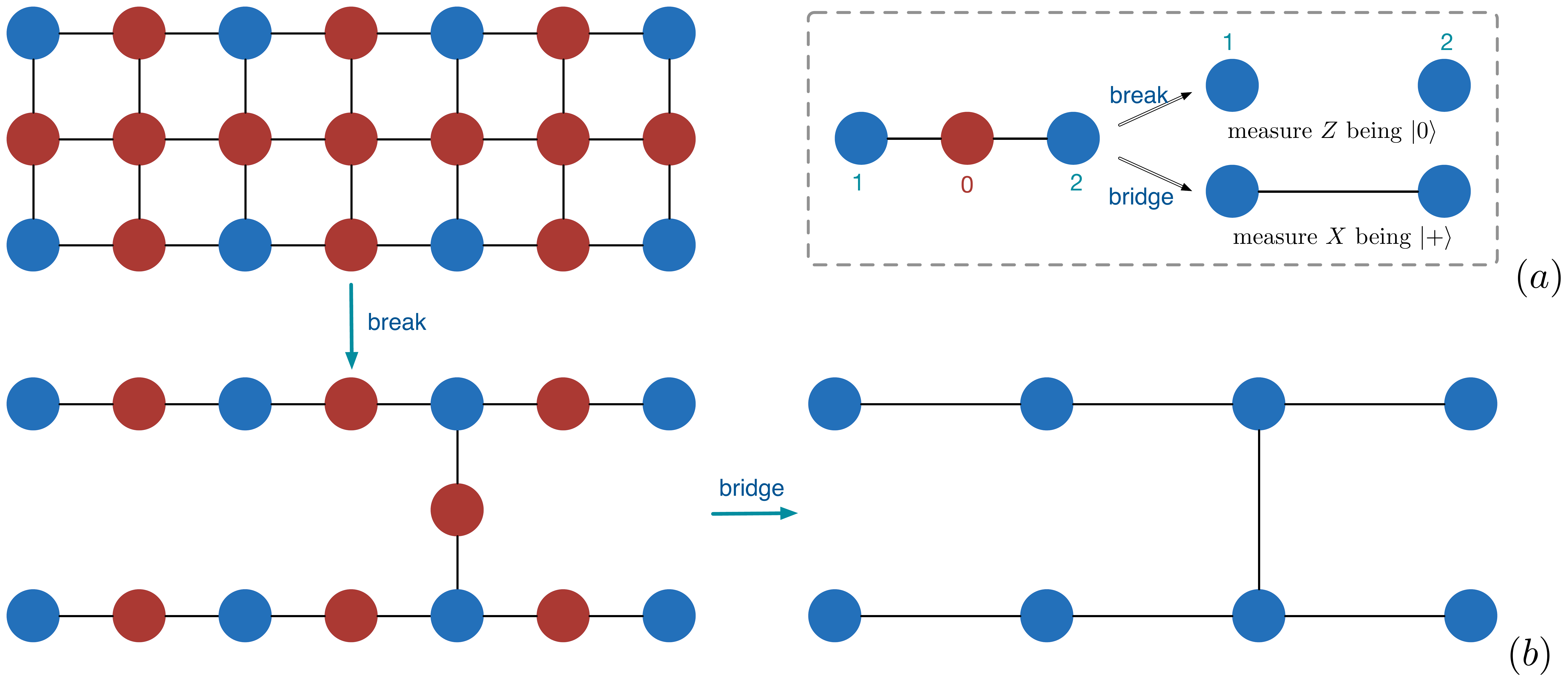}
    \caption{(a) Break and bridge operations. Qubit 0 is first rotated by $R_{z}(\pi/2)$ before measured in $Z$ and $X$ basis respectively to perform the break and bridge operations. (b) Reduce the cluster state to the brickwork state by break and bridge operations. }
    \label{figSup:break_bridge}
\end{figure}

In the main text of our paper, we introduced the ``break" and ``bridge" operations. Here, we include more details of how to reduce a cluster state to a brickwork state by those operations. For the three qubit cluster state in Fig.~\ref{figSup:break_bridge}(a), the red circle is rotated by $R_z(\pi/2)$. The operations acting on qubits 1 and 2 controlled by qubit 0 can be written as 
\begin{equation}\label{eqSup:breakbridge}
\frac{e^{-i \pi/4}}{\sqrt 2}\ket 0_0\otimes I_1\otimes I_2+i\ket 1_0\otimes Z_1\otimes Z_2.
\end{equation}
with an extra global phase. Therefore, by postselecting qubit $0$ being $\ket 0$ by measuring $Z$, we have the operation $I_1\otimes I_2$ on qubits $1$ and $2$, implementing the break operation.  By postselecting qubit $0$ being $\ket+$ by measuring $X$, we have
\begin{equation}
\frac{e^{-i \pi/4}}{\sqrt 2}  \left(I_1\otimes I_2+iZ_1\otimes Z_2 \right)= e^{-i \pi/4} e^{i\pi/4Z_1\otimes Z_2}.
\end{equation}
This is the same as the time evolution of the Ising interaction in the Hamiltonian (Eq.~4 and Eq.~5 of the main text), implementing the bridge operation.

Fig.~\ref{figSup:break_bridge}(b) demonstrates how to convert the cluster state to other graph states such as the brickwork state by the break and bridge operations.

\subsection{Simulation with variation Distance Errors}
This is the most technical part of the computational complex theory in this paper, so we divide it into three parts.

\subsubsection{A \textsf{\#P}-hard problem in worst-case} First of all, we introduce a problem that is \textsf{\#P}-hard in worst-case. Later, we will find that our classically-intractable result for simulating our Ising spin model depends on a conjecture that lifts this problem from worst-case hardness to average-case hardness.

Suppose the probability of measuring result $x=x_1\cdots x_i \cdots x_{m\times n},x_i\in\{0,1\}$ from the quantum sampler is $q_x$ with
\begin{eqnarray}\label{eq:qx_cx}
\nonumber
q_x&=&\left|\bigotimes_i^{m\times n} \bra{+_{x_i}}{e^{-i\mathcal{H}t}}\ket +^{\otimes m\times n}\right|^2\\
&=&\frac{|\bra 0 C_x\ket 0|^2}{2^{mn-m}}
\end{eqnarray}
where $C_x$ is a polynomial size quantum circuit which can be implemented by choosing proper measurement results $x$ and $1/2^{mn-m}$ comes from equal probability for measurement in measurement-based quantum computing. We will show that approximating $q_x$ by $\widetilde q_x$ to the following error
\begin{equation}\label{eq:mix_error1}
|\widetilde q_x-q_x| \le \frac{q_x}{\poly(n)}+\frac{c}{2^{mn}}
\end{equation}
is \textsf{\#P}-hard, where $c$ can be any constant $0\le c<1/2$.

Suppose $f(z)$ is some boolean function which can be computed efficiently by a classical computer. Define 
\begin{equation}
\mbox{gap}(f)\equiv|\{z:f(z)=0\}|-|\{z:f(z)=1\}|=\sum_z(-1)^{f(z)}
\end{equation}
and $\widetilde{\mbox{gap}(f)^2}\equiv 2^{mn}\widetilde{q}_x$. Consider the polynomial size quantum circuit $C_x$ doing the following operation on $\ket 0^{\otimes m}$ ($m=2r$)
\begin{eqnarray}
\nonumber
\mbox{Hadamard gate:\,\,\,\quad}\ket0^{\otimes r}\ket0^{\otimes r}&\Longrightarrow&\ket0^{\otimes m-r}\frac{\sum_z\ket z}{\sqrt{2^r}}\\
\nonumber
\mbox{computing }f(z):\qquad\qquad\qquad
&\Longrightarrow&\ket0^{\otimes r-1}\frac{\sum_z\ket{f(z)}\ket z}{\sqrt{2^r}}\\
\nonumber
\mbox{applying }Z\mbox{ and uncomputing}:\qquad\qquad\qquad
&\Longrightarrow&\ket0^{\otimes r}\frac{\sum_z(-1)^{f(z)}\ket z}{\sqrt{2^r}}\\
\mbox{Hadamard gate}:\qquad\qquad\qquad
&\Longrightarrow&
\ket{0}^{\otimes m}\frac{\sum_z(-1)^{f(z)}}{2^r}+\ket{\mbox{other terms}},
\end{eqnarray}
which means
\begin{equation}
q_x=\frac{|\bra 0 C_x\ket 0|^2}{2^{mn-m}}=\frac{\mbox{gap}(f)^2}{2^{mn}}.
\end{equation}
Thus, Eq.~(\ref{eq:mix_error1}) implies
\begin{equation}
|\widetilde{\mbox{gap}(f)^2}-\mbox{gap}(f)^2|\le\frac{\mbox{gap}(f)^2}{\poly(n)}+c.
\end{equation}
This condition implies $\widetilde{\mbox{gap}(f)^2}$ can estimate $\mbox{gap}(f)^2$ to multiplicative errors since $c<1/2$:
\begin{equation}
|\widetilde{\mbox{gap}(f)^2}-\mbox{gap}(f)^2|\le (c+o(1))\cdot \mbox{gap}(f)^2.
\end{equation}
This is because $\mbox{gap}(f)^2$ is an integer: if $\mbox{gap}(f)^2=0$, then $\widetilde{\mbox{gap}(f)^2}<1/2$ such that we can infer $\mbox{gap}(f)^2=0$, which means $|\widetilde{\mbox{gap}(f)^2}-\mbox{gap}(f)^2|=0$; if $\mbox{gap}(f)^2\ge1$, then $c\le c\cdot\mbox{gap}(f)^2$. Ref.~\cite{PhysRevLett.117.080501} proved that approximating $\mbox{gap}(f)^2$ to multiplicative errors is \textsf{\#P}-hard (actually, they proved that if $f$ is some special boolean function, it is \textsf{GapP}-complete, but this implies the result we need). This proves the worst-case hardness result.

Define the partition function with imaginary temperature $\beta \equiv 1/k_{B}T = i$ as
\begin{equation}
\mathcal{Z}_x=\tr e^{ -i(\mathcal{H}+\sum_ix_i \pi/2 Z_i)}=\sum_{z\in\{+1,-1\}^{mn}}e^{i(\sum_{\left<i,j \right>}\pi/4z_iz_j+\sum_iB^\prime_iz_i)}
\end{equation}
where $B^\prime_i$ depends on $x_i$. Then, 
\begin{eqnarray}
q_x&=&\left|\bigotimes_i^{m\times n} \bra{+_{x_i}}e^{-i\mathcal{H}t}\ket +^{\otimes m\times n}\right|^2\\
&=&\left|\bra{+}^{\otimes m\times n}e^{-i(\mathcal{H}+\sum_ix_i\pi/2 Z_i)t}\ket +^{\otimes m\times n}\right|^2\\
&=&\frac{|\mathcal{Z}_x|^2}{2^{2mn}}
\end{eqnarray}
where $\ket{+_{x}}=Z^x\ket+$ are the bases of $X$.
Restating the above conclusion in terms of the partition function, we get
\begin{theorem}\label{thm:partition_function}
Approximating the partition function to the following error
\begin{equation}
\left|\frac{\widetilde{|\mathcal{Z}_x|^2}}{2^{mn}}-\frac{|\mathcal{Z}_x|^2}{2^{mn}}\right|\le \frac{1}{\mbox{\emph{poly}}(n)}\frac{|\mathcal{Z}_x|^2}{2^{mn}}+c
\end{equation}
is \textsf{\emph{\#P}}-hard in the worst-case, if $0\le c<1/2$. (Notice that the range of $|\mathcal{Z}_x|^2/2^{mn}$ is from 0 to $2^{mn}$ instead of from 0 to 1.)
\end{theorem}

\subsubsection{Classically-intractable for simulation with variation distance error}
The main ingredient is Stockmeyer's theorem \cite{stockmeyer1985approximation} (see Ref.~\cite{aaronson2011computational} or Ref.~\cite{PhysRevLett.117.080501} for  the statement here):
\begin{theorem}
There exists an $\mathsf{FBPP}^\mathsf{NP}$ algorithm which can approximate
\begin{equation}
P=\Pr_x[f(z)=1]=\frac{1}{2^r}\sum_{z\in\{0,1\}^r}f(z)
\end{equation}
by $\widetilde P$, for any boolean function $f:\{0,1\}^r\rightarrow\{0,1\}$, to multiplicative error $|\widetilde P-P|\le P/\mbox{\emph{poly}}(n)$ if $f(z)$ can be computed efficiently given $z$.
\end{theorem}

The probability of any distribution that can be classically efficiently sampled is such kind of $P$: the distribution is produced by tossing the coin and regarding $z$ as the sequence of coin-tossing results, the probability of a specific event is the union of some $z$ such that $f(z)=1$. Hence the above theorem states that any probability in a distribution sampled by a polynomial classical algorithm can be approximated to multiplicative errors in $\mathsf{BPP}^\mathsf{NP}$,  which is contained in the third level of the polynomial hierarchy \cite{stockmeyer1985approximation, aaronson2011computational, PhysRevLett.117.080501}. The probability in the distribution sampled by a quantum algorithm is not $P$ since it involves sums of negative numbers. It can be proved that if $f:\{0,1\}^r\rightarrow\{-1,1\}$, it will still be \textsf{\#P}-hard to approximate the sum to multiplicative errors.

Assume there is a classical sampler that can sample from the distribution $\{p_x\}$. According to Stockmeyer's theorem, $\widetilde p_x$ can be computed in the third level of the polynomial hierarchy such that $|\widetilde p_x-p_x|\le p_x/\poly(n)$. If the distribution $\{p_x\}$ can approximate $\{q_x\}$ to variation distance, i.e., $\sum_x|p_x-q_x|\le\epsilon$. Then $\E_x\left[|p_x-q_x| \right]\le\epsilon/2^{mn}$. Using Markov inequality
\begin{equation}
\Pr_x \left[|p_x-q_x|\ge \frac{\epsilon}{2^{mn}\delta} \right]\le\delta,
\end{equation}
we get
\begin{eqnarray}\label{eq:mix_error2}
\nonumber|\widetilde p_x-q_x| &\le& |\widetilde p_x-p_x|+|p_x-q_x|\\
\nonumber
\mbox{Stockmeyer's theorem:\qquad\qquad}&\le &\frac{p_x}{\poly(n)}+|p_x-q_x|\\
\nonumber
&\le&\frac{q_x+|p_x-q_x|}{\poly(n)}+|p_x-q_x|\\
\nonumber
&=&\frac{q_x}{\poly(n)}+\left(1+\frac{1}{\poly(n)}\right)|p_x-q_x|\\
\nonumber&&\mbox{with }\ge1-\delta \mbox{ fraction of }x\\
\mbox{classically simulable assumption \& Markov inequality:\qquad\qquad}
&\le&\frac{q_x}{\poly(n)}+\frac{\epsilon(1+o(1))}{2^{mn}\delta}.
\end{eqnarray}

We have shown that approximating $q_x$ to a mixture of multiplicative and additive errors in Eq.~(\ref{eq:mix_error2}) is \textsf{\#P}-hard in the worst-case if $\epsilon/\delta<1/2$. Lifting this worst-case hardness result to average-case result, we will get the desired result: If for any $1-\delta$ fraction of instances $x$, approximating $q_x$ to the mixture of the multiplicative and additive errors in Eq.~(\ref{eq:mix_error2}) is still \textsf{\#P}-hard; then if we assume there is a classical sampler that can simulate the distribution of our Ising spin model to variation distance errors, there will exist a $\mathsf{BPP}^\mathsf{NP}$ algorithm that can solve \textsf{\#P}-hard problems, implying the collapse of the polynomial hierarchy.

Restating the above conclusion in terms of the partition function, we get
\begin{theorem}\label{thm:partition_function}
If approximating the partition function to the following error
\begin{equation}\label{eq:partition}
\left|\frac{\widetilde{|\mathcal{Z}_x|^2}}{2^{mn}}-\frac{|\mathcal{Z}_x|^2}{2^{mn}}\right|\le \frac{1}{\mbox{\emph{poly}}(n)}\frac{|\mathcal{Z}_x|^2}{2^{mn}}+
\frac{\epsilon}{\delta}
\end{equation}
is also \textsf{\emph{\#P}}-hard for any $1-\delta$ fraction of instances $x$, then simulating the distribution sampled by our Ising spin model to the variation distance $\epsilon$ is classically intractable, otherwise the polynomial hierarchy will collapse.
\end{theorem}

\subsubsection{Intuition of our average-case hardness conjecture}

Substitute $q_x$ in Eq.~(\ref{eq:mix_error2}) by Eq.~(\ref{eq:qx_cx}) 
\begin{equation}\label{eq:amplitude}
\left|\widetilde{|\bra0C_x\ket0|^2}-|\bra0C_x\ket0|^2\right|\le \frac{|\bra0C_x\ket0|^2}{\poly(n)}+\frac{\epsilon(1+o(1))}{2^{m}\delta}
\end{equation}
where $\widetilde{|\bra0C_x\ket0|^2}$ is an estimation of $|\bra0C_x\ket0|^2$ and $m$ is the width of the circuit $C_x$. The circuit $C_x$ is formed by random 2-qubit gates layer by layer ($n$ layers) similar to Fig.~\ref{figSup:ucircuit}(b). Except some single qubit gates on the boundary, each 2-qubit gate has the form shown in Fig.~\ref{figSup:randomU}, where the angles $\alpha,\beta,\gamma,\delta,\alpha^\prime,\beta^\prime,\gamma^\prime,\delta^\prime$ are chosen from $\{0,\pi/4,\pi/2,3\pi/4,\pi,5\pi/4,3\pi/2,7\pi/4\}$ randomly and independently. This can be verified directly by choosing random measurement results on blue circles in Fig.~1(c) of the main text. If either $\delta$ or $\delta^\prime$ is different from $0$ or $\pi$, this 2-qubit gate will produce entanglement on some product states. In our opinion, with high probability, this kind of circuits will likely produce highly entangled states. Therefore, we conjecture that calculating the amplitudes of the circuit to the error in Eq.~(\ref{eq:amplitude}) is \textsf{\#P}-hard in the average-case.

\begin{figure}[t]
    \includegraphics[width=0.5\linewidth]{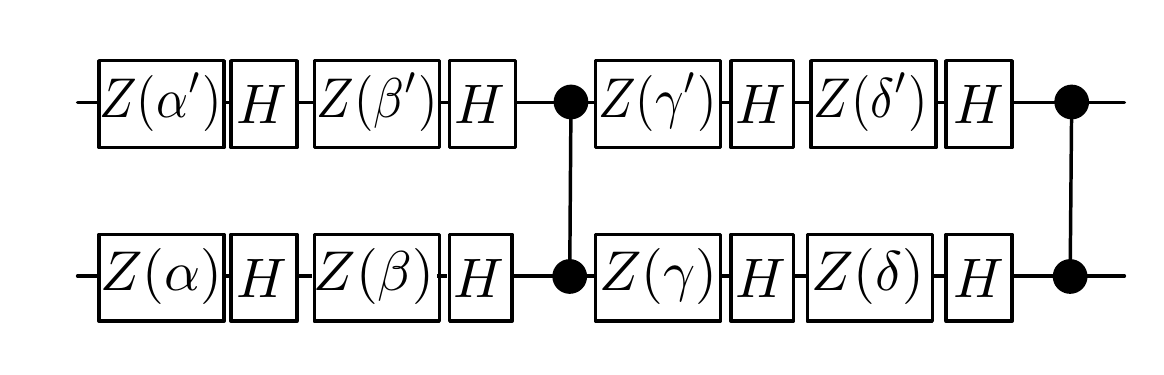}
    \caption{Random 2-qubit gate in $C_x$. $\alpha,\beta,\gamma,\delta,\alpha^\prime,\beta^\prime,\gamma^\prime,\delta^\prime$ are chosen from $\{0,\pi/4,\pi/2,3\pi/4,\pi,5\pi/4,3\pi/2,7\pi/4\}$ randomly and independently.}
    \label{figSup:randomU}
\end{figure}

There is a natural connection between our model and sampling models of random quantum circuits like the one in Ref.~\cite{boixo2016characterizing}. In Ref.~\cite{boixo2016characterizing},   the quantum circuit is basically $\sqrt n$ layers of single qubit gates (chosen from $\{X^{1/2},Y^{1/2},R_z(\pi/4)\}$ randomly) and control-$Z$ gates applied to $\sqrt n\times \sqrt n$ input qubits on square lattice. The intuition of classical hardness of this sampling problem is from the relation between quantum chaos and random quantum circuits. The distribution produced by their sampling model is expected to satisfy the Porter-Thomas distribution \cite{PhysRev.104.483} with a sufficient circuit depth. This is supported by numerical simulations in Ref.~\cite{boixo2016characterizing}. Then there is a large fraction of $|\bra 0U\ket z|^2\ge 1/2^m$ where $U$ is a random circuit, which implies that approximating output probabilities to multiplicative errors is \textsf{\#P}-hard in average-case and the noncollapse of the polynomial hierarchy is sufficient to prove the classical hardness result. Although the ensembles used in our model and the one in Ref.~\cite{boixo2016characterizing} are different, we think there is no fundamental difference since they both try to produce sufficiently random quantum circuits. Besides, it is expected that the distribution of our model approaches the Porter-Thomas distribution if $n \sim m$ because the ``input" in our model is on a linear array (the depth is expected to grow as $n^{1/D}$ for a $D$ dimensional qubit lattice. See corresponding discussions in Ref.~\cite{boixo2016characterizing}). Therefore, we should be able to convert Eq.~(\ref{eq:partition}) and Eq.~(\ref{eq:amplitude}) with multiplicative errors in our conjecture to be similar to the one in Ref.~\cite{PhysRevLett.117.080501}.

\subsection{Certification to variation Distance Errors}

With the reasonable assumption that the errors of $X$ measurements are local and small (scales as $O(1/mn)$), we can certify whether the distribution sampled by a quantum sampler in the laboratory satisfies the variation distance bound. First, we give the condition that the measurement errors should satisfy; then we reduce the certification to bounding the trace distance between the ideal final state and the actual one prepared in the laboratory (before measurement). 

Suppose $\{q_x^\prime\}$ is the distribution sampled by the quantum sampler and the density matrix just before measurements is $\rho^\prime$; $\{q_x\}$ is the ideal one with the corresponding density matrix $\rho$. Denote the trace distance by $D(\rho,\rho^\prime)= \tr(|\rho-\rho^{\prime}|)/2$. It is known that  \cite{nielsen2010quantum} 
\begin{equation}
\sum_x|q_x-q^\prime_x|\le D(\rho,\rho^\prime).
\end{equation}
So if the measurements are perfect, bounding the trace distance will imply that variation distance is bounded. 

Let us consider measurement imperfections. Denote the ideal measurement as a quantum operator $\mathcal E$ and the imperfect one as $\mathcal E^\prime$. If the measurement errors are small and local, $\mathcal E^\prime$ can be approximated as
\begin{eqnarray}
\mathcal E^\prime\approx \mathcal E\circ \left(\mathcal I+\varepsilon \sum_i\omega_i\right)
\end{eqnarray}
where $\mathcal I$ is the identity quantum operation, $\omega_i$ is some local operation around spin $i$, and $\varepsilon$ is some small number. Bounding the variation distance can be reduced by 
\begin{equation}
\sum_x|q_x-q^\prime_x|=D(\mathcal E(\rho),\mathcal E^\prime(\rho^\prime))\le D(\mathcal E(\rho),\mathcal E(\rho^\prime))+D(\mathcal E(\rho^\prime),\mathcal E^\prime(\rho^\prime))\le D(\rho,\rho^\prime)+D(\mathcal E(\rho^\prime),\mathcal E^\prime(\rho^\prime)).
\end{equation}
The term $D(\rho,\rho^\prime)$ characterizes the error produced in the process of preparing the final state (time evolution and initial state preparation errors). The term $D(\mathcal E(\rho^\prime),\mathcal E^\prime(\rho^\prime))$ characterizes the error due to imperfect measurements.

We divide the certification of the variation distance error into two parts:
\begin{eqnarray}
\nonumber
D(\rho,\rho^\prime)&\le& \epsilon_d\\
\nonumber
D(\mathcal E(\rho^\prime),\mathcal E^\prime(\rho^\prime))&\le&\epsilon_m\\
\epsilon_d+\epsilon_m&\le &\epsilon.
\end{eqnarray}
The error due to imperfect measurements is
\begin{equation}
\|\mathcal E^\prime(\sigma)-\mathcal E(\sigma)\|\approx\left\|\varepsilon \mathcal E\circ\left(\sum_i\omega_i\right)
(\sigma)\right\|\le mn\varepsilon
\end{equation}
where $\sigma$ is some arbitrary density matrix.
So as long as the measurement error on every spin can be made smaller than $\varepsilon=\epsilon_m/(mn)$, it can be guaranteed that the total measurement error is bounded by $\epsilon_m$.

The remaining is to certify whether $D(\rho,\rho^\prime)\le\epsilon_d$. We reduce the problem to certifying whether the state produced in the laboratory is close to the ideal state, which is made to be the ground state of a given local gapped Hamiltonian. The method in Ref.~\cite{hangleiter2016direct} can achieve this task. Recall a lemma in Ref.~\cite{hangleiter2016direct}: 
\begin{lemma}
Suppose $\rho$ is the ground state of $H=\sum_\lambda h_\lambda$ where $h_\lambda$ is a local Hermitian operator, the ground state is unique and the ground state energy is 0. To estimate $\tr(h_\lambda \rho^\prime)$ where $\rho^\prime$ is the state produced in the laboratory, $M$ measurements on $\rho^\prime$ in the basis of $h_\lambda$ are needed. By summing over all the estimations of $h_\lambda$, we can get an estimation of $\tr(H\rho^\prime)$. By this estimation, we can estimate $F(\rho,\rho^\prime)= \tr (\rho \rho^{\prime}) $ by $F^*$ where 
\begin{equation}
\Pr[|F^*-F|\le\epsilon^\prime]\ge1-\alpha.
\end{equation}
If we choose $M$ as
\begin{equation}
M\ge \frac{Jm^2n^2}{2\Delta^2\epsilon^{\prime2}}\ln\left[-\frac{mn+1}{\ln(1-\alpha)}\right]\approx
\frac{Jm^2n^2}{2\Delta^2\epsilon^{\prime2}}\left(\ln mn+\ln\frac{1}{\alpha}\right)\mbox{ for }m,n\mbox{ large and }\alpha\mbox{ small}
\end{equation}
where $\Delta$ is the energy gap and $J=\max_\lambda\|h_\lambda\|$.
\end{lemma}
Because $D(\rho,\rho^\prime)\le\sqrt{1-F^2(\rho,\rho^\prime)}$, $F(\rho,\rho^\prime)\ge\sqrt{1-\epsilon_d^2}$ implies $D(\rho,\rho^\prime)\le \epsilon_d$. So we require
\begin{equation}
F^*\ge\sqrt{1-\epsilon_d^2}+\epsilon^\prime.
\end{equation}

In our problem, the Hamiltonian is
\begin{equation}
H^\prime_{ \text{brickwork} } = \dfrac{1}{2} \sum_{i}
\left( I-R_{z}(\theta_i) X_i R_{z}^\dag(\theta_i)\textstyle\prod_{j\in \text{neighbor of} \,i}Z_j \right)
\end{equation}
on the brickwork lattice as shown in Fig.~1 of the main text, and $J=1$, $\Delta=1$.

If we choose $\epsilon_d=O(\epsilon)$, $\epsilon_m=O(\epsilon)$ and $\epsilon^\prime=O(\epsilon^2)$, then we need to measure each local term in the Hamiltonian $M=O(m^2n^2r/\epsilon^4)$ times to get a confidence level of $1-2^{-O(r)}$. The certification protocol is therefore efficient.

\subsection{Hardness of Classically Simulating the Square Lattice Model to variation Distance Errors}

When doing break and bridge operations, we need to measure $Z$ being $\ket 0$ and $X$ being $\ket +$ on the red circles in Fig.~\ref{figSup:break_bridge}, but the results $\ket 1$ and $\ket -$ are also present as we sample. According to Eq.~(\ref{eqSup:breakbridge}), we can conclude
\begin{itemize}
\item Measuring $Z$ on qubit 0, the probabilities of getting $\ket 0$ and $\ket 1$ are both $1/2$. When the result is $\ket 1$, the operation is $iZ_1\otimes Z_2$, so the effect is just flipping the measurement result on the blue circles in Fig.~\ref{figSup:break_bridge}.
\item Measuring $X$ on qubit 0, the probability of getting $\ket +$ and $\ket-$ are also $1/2$ each. When the result is $\ket -$, the operation is
\begin{equation}
\frac{e^{-i \pi/4}}{\sqrt 2}  \left(I_1\otimes I_2-iZ_1\otimes Z_2 \right)= e^{-i \pi/4} e^{-i\pi/4Z_1\otimes Z_2}.
\end{equation}
Since
\begin{equation}
e^{-i\pi/4Z_1\otimes Z_2}=-ie^{i\pi/4Z_1\otimes Z_2}Z_1\otimes Z_2,
\end{equation}
the effect is also flipping the measurement result on the blue circles.
\end{itemize}
Denote the measurement result on blue circles as $x^\prime$ and result on red circles as $y$ (for the bridge operation, denote $\ket+$ as 0) and $q_x$ is the probability of measuring $x$ on the brickwork model. Because the effect of $y$ may be just flipping some bit of $x$, given $y$, we can infer $x$ and $x^\prime$ from each other. Besides, $q_y\equiv\sum_{x^\prime}q_{x^\prime,y}=1/2^r$ where $r$ is the number of red circles (actually, $r=3mn-2m-2n+1$) and
$q_{x^\prime|y}=q_x$, so
\begin{equation}\label{eqSup:qx}
\sum_y q_{x^\prime,y}=\sum_yq_{x^\prime|y}q_y=\sum_y\frac{1}{2^r}q_x=q_x.
\end{equation}

Suppose there exists a quantum sampler that can generate a distribution $\{p_{x^\prime,y}\}$ to approximate the distribution of square lattice model to variation distance errors:
\begin{equation}
\sum_{x^\prime,y}|p_{x^\prime,y}-q_{x^\prime,y}|\le\epsilon.
\end{equation}
We can then define a new classical sampler to simulate the distribution of the brickwork model: suppose the outcome is $x^\prime,y$ and define the result to be $x$ ($x^\prime,y$ can determine a unique $x$), so the probability of getting $x$ is $p_x=\sum_yp_{x^\prime,y}$, implying
\begin{eqnarray}
\nonumber
\sum_{x^\prime,y}|p_{x^\prime,y}-q_{x^\prime,y}|&=&\sum_{x,y}|p_{x^\prime,y}-q_{x^\prime,y}|\\
\nonumber
&\ge&\sum_x|\sum_yp_{x^\prime,y}-\sum_yq_{x^\prime,y}|\\
&=&\sum_x|p_x-q_x|.
\end{eqnarray}
The first equality is because given $y$, $x$ and $x^\prime$ can determine each other. The last equality is due to the definition of $p_x$ and Eq.~(\ref{eqSup:qx}). This implies that there exists a classical sampler to simulate the brickwork model. So the hardness result of the square lattice model is based on the same conjectures (polynomial hierarchy does not collapse and Theorem~\ref{thm:partition_function}).

\end{widetext}


\begin{thebibliography}{46}%
\makeatletter
\providecommand \@ifxundefined [1]{%
 \@ifx{#1\undefined}
}%
\providecommand \@ifnum [1]{%
 \ifnum #1\expandafter \@firstoftwo
 \else \expandafter \@secondoftwo
 \fi
}%
\providecommand \@ifx [1]{%
 \ifx #1\expandafter \@firstoftwo
 \else \expandafter \@secondoftwo
 \fi
}%
\providecommand \natexlab [1]{#1}%
\providecommand \enquote  [1]{``#1''}%
\providecommand \bibnamefont  [1]{#1}%
\providecommand \bibfnamefont [1]{#1}%
\providecommand \citenamefont [1]{#1}%
\providecommand \href@noop [0]{\@secondoftwo}%
\providecommand \href [0]{\begingroup \@sanitize@url \@href}%
\providecommand \@href[1]{\@@startlink{#1}\@@href}%
\providecommand \@@href[1]{\endgroup#1\@@endlink}%
\providecommand \@sanitize@url [0]{\catcode `\\12\catcode `\$12\catcode
  `\&12\catcode `\#12\catcode `\^12\catcode `\_12\catcode `\%12\relax}%
\providecommand \@@startlink[1]{}%
\providecommand \@@endlink[0]{}%
\providecommand \url  [0]{\begingroup\@sanitize@url \@url }%
\providecommand \@url [1]{\endgroup\@href {#1}{\urlprefix }}%
\providecommand \urlprefix  [0]{URL }%
\providecommand \Eprint [0]{\href }%
\providecommand \doibase [0]{http://dx.doi.org/}%
\providecommand \selectlanguage [0]{\@gobble}%
\providecommand \bibinfo  [0]{\@secondoftwo}%
\providecommand \bibfield  [0]{\@secondoftwo}%
\providecommand \translation [1]{[#1]}%
\providecommand \BibitemOpen [0]{}%
\providecommand \bibitemStop [0]{}%
\providecommand \bibitemNoStop [0]{.\EOS\space}%
\providecommand \EOS [0]{\spacefactor3000\relax}%
\providecommand \BibitemShut  [1]{\csname bibitem#1\endcsname}%
\let\auto@bib@innerbib\@empty
\bibitem [{\citenamefont {Nielsen}\ and\ \citenamefont
  {Chuang}(2010)}]{nielsen2010quantum}%
  \BibitemOpen
  \bibfield  {author} {\bibinfo {author} {\bibfnamefont {M.~A.}\ \bibnamefont
  {Nielsen}}\ and\ \bibinfo {author} {\bibfnamefont {I.~L.}\ \bibnamefont
  {Chuang}},\ }\href@noop {} {\emph {\bibinfo {title} {Quantum computation and
  quantum information}}}\ (\bibinfo  {publisher} {Cambridge university press},\
  \bibinfo {year} {2010})\BibitemShut {NoStop}%
\bibitem [{\citenamefont {Ladd}\ \emph {et~al.}(2010)\citenamefont {Ladd},
  \citenamefont {Jelezko}, \citenamefont {Laflamme}, \citenamefont {Nakamura},
  \citenamefont {Monroe},\ and\ \citenamefont {O'Brien}}]{Ladd2010Quantum}%
  \BibitemOpen
  \bibfield  {author} {\bibinfo {author} {\bibfnamefont {T.~D.}\ \bibnamefont
  {Ladd}}, \bibinfo {author} {\bibfnamefont {F.}~\bibnamefont {Jelezko}},
  \bibinfo {author} {\bibfnamefont {R.}~\bibnamefont {Laflamme}}, \bibinfo
  {author} {\bibfnamefont {Y.}~\bibnamefont {Nakamura}}, \bibinfo {author}
  {\bibfnamefont {C.}~\bibnamefont {Monroe}}, \ and\ \bibinfo {author}
  {\bibfnamefont {J.~L.}\ \bibnamefont {O'Brien}},\ }\href
  {http://dx.doi.org/10.1038/nature08812} {\bibfield  {journal} {\bibinfo
  {journal} {Nature}\ }\textbf {\bibinfo {volume} {464}},\ \bibinfo {pages}
  {45} (\bibinfo {year} {2010})}\BibitemShut {NoStop}%
\bibitem [{\citenamefont {Shor}(1994)}]{shor1994algorithms}%
  \BibitemOpen
  \bibfield  {author} {\bibinfo {author} {\bibfnamefont {P.~W.}\ \bibnamefont
  {Shor}},\ }in\ \href {\doibase 10.1109/SFCS.1994.365700} {\emph {\bibinfo
  {booktitle} {Foundations of Computer Science, 1994 Proceedings., 35th Annual
  Symposium on}}}\ (\bibinfo {year} {1994})\ pp.\ \bibinfo {pages}
  {124--134}\BibitemShut {NoStop}%
\bibitem [{\citenamefont {Harrow}\ \emph {et~al.}(2009)\citenamefont {Harrow},
  \citenamefont {Hassidim},\ and\ \citenamefont {Lloyd}}]{harrow2009quantum}%
  \BibitemOpen
  \bibfield  {author} {\bibinfo {author} {\bibfnamefont {A.~W.}\ \bibnamefont
  {Harrow}}, \bibinfo {author} {\bibfnamefont {A.}~\bibnamefont {Hassidim}}, \
  and\ \bibinfo {author} {\bibfnamefont {S.}~\bibnamefont {Lloyd}},\ }\href
  {\doibase 10.1103/PhysRevLett.103.150502} {\bibfield  {journal} {\bibinfo
  {journal} {Phys. Rev. Lett.}\ }\textbf {\bibinfo {volume} {103}},\ \bibinfo
  {pages} {150502} (\bibinfo {year} {2009})}\BibitemShut {NoStop}%
\bibitem [{\citenamefont {Aaronson}\ and\ \citenamefont
  {Arkhipov}(2011)}]{aaronson2011computational}%
  \BibitemOpen
  \bibfield  {author} {\bibinfo {author} {\bibfnamefont {S.}~\bibnamefont
  {Aaronson}}\ and\ \bibinfo {author} {\bibfnamefont {A.}~\bibnamefont
  {Arkhipov}},\ }in\ \href {\doibase 10.1145/1993636.1993682} {\emph {\bibinfo
  {booktitle} {Proceedings of the Forty-third Annual ACM Symposium on Theory of
  Computing}}},\ \bibinfo {series and number} {STOC '11}\ (\bibinfo
  {publisher} {ACM},\ \bibinfo {address} {New York, NY, USA},\ \bibinfo {year}
  {2011})\ pp.\ \bibinfo {pages} {333--342}\BibitemShut {NoStop}%
\bibitem [{\citenamefont {Bremner}\ \emph {et~al.}(2010)\citenamefont
  {Bremner}, \citenamefont {Jozsa},\ and\ \citenamefont
  {Shepherd}}]{bremner2011classical}%
  \BibitemOpen
  \bibfield  {author} {\bibinfo {author} {\bibfnamefont {M.~J.}\ \bibnamefont
  {Bremner}}, \bibinfo {author} {\bibfnamefont {R.}~\bibnamefont {Jozsa}}, \
  and\ \bibinfo {author} {\bibfnamefont {D.~J.}\ \bibnamefont {Shepherd}},\
  }\href {\doibase 10.1098/rspa.2010.0301} {\bibfield  {journal} {\bibinfo
  {journal} {Proc. R. Soc. A}\ }\textbf {\bibinfo {volume} {467}},\ \bibinfo
  {pages} {459} (\bibinfo {year} {2010})}\BibitemShut {NoStop}%
\bibitem [{\citenamefont {Bremner}\ \emph
  {et~al.}(2016{\natexlab{a}})\citenamefont {Bremner}, \citenamefont
  {Montanaro},\ and\ \citenamefont {Shepherd}}]{PhysRevLett.117.080501}%
  \BibitemOpen
  \bibfield  {author} {\bibinfo {author} {\bibfnamefont {M.~J.}\ \bibnamefont
  {Bremner}}, \bibinfo {author} {\bibfnamefont {A.}~\bibnamefont {Montanaro}},
  \ and\ \bibinfo {author} {\bibfnamefont {D.~J.}\ \bibnamefont {Shepherd}},\
  }\href {\doibase 10.1103/PhysRevLett.117.080501} {\bibfield  {journal}
  {\bibinfo  {journal} {Phys. Rev. Lett.}\ }\textbf {\bibinfo {volume} {117}},\
  \bibinfo {pages} {080501} (\bibinfo {year} {2016}{\natexlab{a}})}\BibitemShut
  {NoStop}%
\bibitem [{\citenamefont {Fujii}\ and\ \citenamefont
  {Tamate}(2016)}]{Fujii2016Computational}%
  \BibitemOpen
  \bibfield  {author} {\bibinfo {author} {\bibfnamefont {K.}~\bibnamefont
  {Fujii}}\ and\ \bibinfo {author} {\bibfnamefont {S.}~\bibnamefont {Tamate}},\
  }\href {\doibase 10.1038/srep25598} {\bibfield  {journal} {\bibinfo
  {journal} {Sci. Rep.}\ }\textbf {\bibinfo {volume} {6}},\ \bibinfo {pages}
  {25598} (\bibinfo {year} {2016})}\BibitemShut {NoStop}%
\bibitem [{\citenamefont {Bremner}\ \emph
  {et~al.}(2016{\natexlab{b}})\citenamefont {Bremner}, \citenamefont
  {Montanaro},\ and\ \citenamefont {Shepherd}}]{bremner2016achieving}%
  \BibitemOpen
  \bibfield  {author} {\bibinfo {author} {\bibfnamefont {M.~J.}\ \bibnamefont
  {Bremner}}, \bibinfo {author} {\bibfnamefont {A.}~\bibnamefont {Montanaro}},
  \ and\ \bibinfo {author} {\bibfnamefont {D.~J.}\ \bibnamefont {Shepherd}},\
  }\href@noop {} {\bibfield  {journal} {\bibinfo  {journal} {ArXiv e-prints}\ }
  (\bibinfo {year} {2016}{\natexlab{b}})},\ \Eprint
  {http://arxiv.org/abs/1610.01808} {arXiv:1610.01808 [quant-ph]} \BibitemShut
  {NoStop}%
\bibitem [{\citenamefont {Morimae}\ \emph {et~al.}(2014)\citenamefont
  {Morimae}, \citenamefont {Fujii},\ and\ \citenamefont
  {Fitzsimons}}]{morimae2014hardness}%
  \BibitemOpen
  \bibfield  {author} {\bibinfo {author} {\bibfnamefont {T.}~\bibnamefont
  {Morimae}}, \bibinfo {author} {\bibfnamefont {K.}~\bibnamefont {Fujii}}, \
  and\ \bibinfo {author} {\bibfnamefont {J.~F.}\ \bibnamefont {Fitzsimons}},\
  }\href {\doibase 10.1103/PhysRevLett.112.130502} {\bibfield  {journal}
  {\bibinfo  {journal} {Phys. Rev. Lett.}\ }\textbf {\bibinfo {volume} {112}},\
  \bibinfo {pages} {130502} (\bibinfo {year} {2014})}\BibitemShut {NoStop}%
\bibitem [{\citenamefont {{Fujii}}\ \emph {et~al.}(2014)\citenamefont
  {{Fujii}}, \citenamefont {{Kobayashi}}, \citenamefont {{Morimae}},
  \citenamefont {{Nishimura}}, \citenamefont {{Tamate}},\ and\ \citenamefont
  {{Tani}}}]{fujii2014impossibility}%
  \BibitemOpen
  \bibfield  {author} {\bibinfo {author} {\bibfnamefont {K.}~\bibnamefont
  {{Fujii}}}, \bibinfo {author} {\bibfnamefont {H.}~\bibnamefont
  {{Kobayashi}}}, \bibinfo {author} {\bibfnamefont {T.}~\bibnamefont
  {{Morimae}}}, \bibinfo {author} {\bibfnamefont {H.}~\bibnamefont
  {{Nishimura}}}, \bibinfo {author} {\bibfnamefont {S.}~\bibnamefont
  {{Tamate}}}, \ and\ \bibinfo {author} {\bibfnamefont {S.}~\bibnamefont
  {{Tani}}},\ }\href@noop {} {\bibfield  {journal} {\bibinfo  {journal} {ArXiv
  e-prints}\ } (\bibinfo {year} {2014})},\ \Eprint
  {http://arxiv.org/abs/1409.6777} {arXiv:1409.6777 [quant-ph]} \BibitemShut
  {NoStop}%
\bibitem [{\citenamefont {{Bouland}}\ \emph {et~al.}(2016)\citenamefont
  {{Bouland}}, \citenamefont {{Man{\v c}inska}},\ and\ \citenamefont
  {{Zhang}}}]{bouland2016complexity}%
  \BibitemOpen
  \bibfield  {author} {\bibinfo {author} {\bibfnamefont {A.}~\bibnamefont
  {{Bouland}}}, \bibinfo {author} {\bibfnamefont {L.}~\bibnamefont {{Man{\v
  c}inska}}}, \ and\ \bibinfo {author} {\bibfnamefont {X.}~\bibnamefont
  {{Zhang}}},\ }\href@noop {} {\bibfield  {journal} {\bibinfo  {journal} {ArXiv
  e-prints}\ } (\bibinfo {year} {2016})},\ \Eprint
  {http://arxiv.org/abs/1602.04145} {arXiv:1602.04145 [quant-ph]} \BibitemShut
  {NoStop}%
\bibitem [{\citenamefont {{Farhi}}\ and\ \citenamefont
  {{Harrow}}(2016)}]{2016arXiv160207674F}%
  \BibitemOpen
  \bibfield  {author} {\bibinfo {author} {\bibfnamefont {E.}~\bibnamefont
  {{Farhi}}}\ and\ \bibinfo {author} {\bibfnamefont {A.~W.}\ \bibnamefont
  {{Harrow}}},\ }\href@noop {} {\bibfield  {journal} {\bibinfo  {journal}
  {ArXiv e-prints}\ } (\bibinfo {year} {2016})},\ \Eprint
  {http://arxiv.org/abs/1602.07674} {arXiv:1602.07674 [quant-ph]} \BibitemShut
  {NoStop}%
\bibitem [{\citenamefont {Boixo}\ \emph {et~al.}(2016)\citenamefont {Boixo},
  \citenamefont {Isakov}, \citenamefont {Smelyanskiy}, \citenamefont {Babbush},
  \citenamefont {Ding}, \citenamefont {Jiang}, \citenamefont {Martinis},\ and\
  \citenamefont {Neven}}]{boixo2016characterizing}%
  \BibitemOpen
  \bibfield  {author} {\bibinfo {author} {\bibfnamefont {S.}~\bibnamefont
  {Boixo}}, \bibinfo {author} {\bibfnamefont {S.~V.}\ \bibnamefont {Isakov}},
  \bibinfo {author} {\bibfnamefont {V.~N.}\ \bibnamefont {Smelyanskiy}},
  \bibinfo {author} {\bibfnamefont {R.}~\bibnamefont {Babbush}}, \bibinfo
  {author} {\bibfnamefont {N.}~\bibnamefont {Ding}}, \bibinfo {author}
  {\bibfnamefont {Z.}~\bibnamefont {Jiang}}, \bibinfo {author} {\bibfnamefont
  {J.~M.}\ \bibnamefont {Martinis}}, \ and\ \bibinfo {author} {\bibfnamefont
  {H.}~\bibnamefont {Neven}},\ }\href@noop {} {\bibfield  {journal} {\bibinfo
  {journal} {ArXiv e-prints}\ } (\bibinfo {year} {2016})},\ \Eprint
  {http://arxiv.org/abs/1608.00263} {arXiv:1608.00263 [quant-ph]} \BibitemShut
  {NoStop}%
\bibitem [{\citenamefont {{Fujii}}(2016)}]{fujii2016noise}%
  \BibitemOpen
  \bibfield  {author} {\bibinfo {author} {\bibfnamefont {K.}~\bibnamefont
  {{Fujii}}},\ }\href@noop {} {\bibfield  {journal} {\bibinfo  {journal} {ArXiv
  e-prints}\ } (\bibinfo {year} {2016})},\ \Eprint
  {http://arxiv.org/abs/1610.03632} {arXiv:1610.03632 [quant-ph]} \BibitemShut
  {NoStop}%
\bibitem [{\citenamefont {Arora}\ and\ \citenamefont
  {Barak}(2009)}]{arora2009computational}%
  \BibitemOpen
  \bibfield  {author} {\bibinfo {author} {\bibfnamefont {S.}~\bibnamefont
  {Arora}}\ and\ \bibinfo {author} {\bibfnamefont {B.}~\bibnamefont {Barak}},\
  }\href@noop {} {\emph {\bibinfo {title} {Computational complexity: a modern
  approach}}}\ (\bibinfo  {publisher} {Cambridge University Press},\ \bibinfo
  {year} {2009})\BibitemShut {NoStop}%
\bibitem [{sm()}]{sm}%
  \BibitemOpen
  \href@noop {} {}\bibinfo {note} {See Supplemental Material [url], which
  includes Refs.\cite{stockmeyer1985approximation,PhysRev.104.483}, for more
  details on the proof of the hardness result and the simulation and
  certification with variation distance errors.}\BibitemShut {Stop}%
\bibitem [{\citenamefont {Stockmeyer}(1985)}]{stockmeyer1985approximation}%
  \BibitemOpen
  \bibfield  {author} {\bibinfo {author} {\bibfnamefont {L.}~\bibnamefont
  {Stockmeyer}},\ }\href {\doibase 10.1137/0214060} {\bibfield  {journal}
  {\bibinfo  {journal} {SIAM J. Comput.}\ }\textbf {\bibinfo {volume} {14}},\
  \bibinfo {pages} {849} (\bibinfo {year} {1985})}\BibitemShut {NoStop}%
\bibitem [{\citenamefont {Porter}\ and\ \citenamefont
  {Thomas}(1956)}]{PhysRev.104.483}%
  \BibitemOpen
  \bibfield  {author} {\bibinfo {author} {\bibfnamefont {C.~E.}\ \bibnamefont
  {Porter}}\ and\ \bibinfo {author} {\bibfnamefont {R.~G.}\ \bibnamefont
  {Thomas}},\ }\href {\doibase 10.1103/PhysRev.104.483} {\bibfield  {journal}
  {\bibinfo  {journal} {Phys. Rev.}\ }\textbf {\bibinfo {volume} {104}},\
  \bibinfo {pages} {483} (\bibinfo {year} {1956})}\BibitemShut {NoStop}%
\bibitem [{\citenamefont {Broome}\ \emph {et~al.}(2013)\citenamefont {Broome},
  \citenamefont {Fedrizzi}, \citenamefont {Rahimi-Keshari}, \citenamefont
  {Dove}, \citenamefont {Aaronson}, \citenamefont {Ralph},\ and\ \citenamefont
  {White}}]{Broome2013Photonic}%
  \BibitemOpen
  \bibfield  {author} {\bibinfo {author} {\bibfnamefont {M.~A.}\ \bibnamefont
  {Broome}}, \bibinfo {author} {\bibfnamefont {A.}~\bibnamefont {Fedrizzi}},
  \bibinfo {author} {\bibfnamefont {S.}~\bibnamefont {Rahimi-Keshari}},
  \bibinfo {author} {\bibfnamefont {J.}~\bibnamefont {Dove}}, \bibinfo {author}
  {\bibfnamefont {S.}~\bibnamefont {Aaronson}}, \bibinfo {author}
  {\bibfnamefont {T.~C.}\ \bibnamefont {Ralph}}, \ and\ \bibinfo {author}
  {\bibfnamefont {A.~G.}\ \bibnamefont {White}},\ }\href {\doibase
  10.1126/science.1231440} {\bibfield  {journal} {\bibinfo  {journal}
  {Science}\ }\textbf {\bibinfo {volume} {339}},\ \bibinfo {pages} {794}
  (\bibinfo {year} {2013})}\BibitemShut {NoStop}%
\bibitem [{\citenamefont {Spring}\ \emph {et~al.}(2013)\citenamefont {Spring},
  \citenamefont {Metcalf}, \citenamefont {Humphreys}, \citenamefont
  {Kolthammer}, \citenamefont {Jin}, \citenamefont {Barbieri}, \citenamefont
  {Datta}, \citenamefont {Thomas-Peter}, \citenamefont {Langford},
  \citenamefont {Kundys}, \citenamefont {Gates}, \citenamefont {Smith},
  \citenamefont {Smith},\ and\ \citenamefont {Walmsley}}]{Spring2013Boson}%
  \BibitemOpen
  \bibfield  {author} {\bibinfo {author} {\bibfnamefont {J.~B.}\ \bibnamefont
  {Spring}}, \bibinfo {author} {\bibfnamefont {B.~J.}\ \bibnamefont {Metcalf}},
  \bibinfo {author} {\bibfnamefont {P.~C.}\ \bibnamefont {Humphreys}}, \bibinfo
  {author} {\bibfnamefont {W.~S.}\ \bibnamefont {Kolthammer}}, \bibinfo
  {author} {\bibfnamefont {X.-M.}\ \bibnamefont {Jin}}, \bibinfo {author}
  {\bibfnamefont {M.}~\bibnamefont {Barbieri}}, \bibinfo {author}
  {\bibfnamefont {A.}~\bibnamefont {Datta}}, \bibinfo {author} {\bibfnamefont
  {N.}~\bibnamefont {Thomas-Peter}}, \bibinfo {author} {\bibfnamefont {N.~K.}\
  \bibnamefont {Langford}}, \bibinfo {author} {\bibfnamefont {D.}~\bibnamefont
  {Kundys}}, \bibinfo {author} {\bibfnamefont {J.~C.}\ \bibnamefont {Gates}},
  \bibinfo {author} {\bibfnamefont {B.~J.}\ \bibnamefont {Smith}}, \bibinfo
  {author} {\bibfnamefont {P.~G.~R.}\ \bibnamefont {Smith}}, \ and\ \bibinfo
  {author} {\bibfnamefont {I.~A.}\ \bibnamefont {Walmsley}},\ }\href {\doibase
  10.1126/science.1231692} {\bibfield  {journal} {\bibinfo  {journal}
  {Science}\ }\textbf {\bibinfo {volume} {339}},\ \bibinfo {pages} {798}
  (\bibinfo {year} {2013})}\BibitemShut {NoStop}%
\bibitem [{\citenamefont {Feynman}(1982)}]{feynman1982simulating}%
  \BibitemOpen
  \bibfield  {author} {\bibinfo {author} {\bibfnamefont {R.~P.}\ \bibnamefont
  {Feynman}},\ }\href {\doibase 10.1007/BF02650179} {\bibfield  {journal}
  {\bibinfo  {journal} {Int. J. Theoret. Phys.}\ }\textbf {\bibinfo {volume}
  {21}},\ \bibinfo {pages} {467} (\bibinfo {year} {1982})}\BibitemShut
  {NoStop}%
\bibitem [{\citenamefont {Lloyd}(1996)}]{lloyd1996universal}%
  \BibitemOpen
  \bibfield  {author} {\bibinfo {author} {\bibfnamefont {S.}~\bibnamefont
  {Lloyd}},\ }\href
  {http://search.proquest.com/docview/213562780?accountid=14426} {\bibfield
  {journal} {\bibinfo  {journal} {Science}\ }\textbf {\bibinfo {volume}
  {273}},\ \bibinfo {pages} {1073} (\bibinfo {year} {1996})}\BibitemShut
  {NoStop}%
\bibitem [{\citenamefont {Buluta}\ and\ \citenamefont
  {Nori}(2009)}]{Buluta2009Quantum}%
  \BibitemOpen
  \bibfield  {author} {\bibinfo {author} {\bibfnamefont {I.}~\bibnamefont
  {Buluta}}\ and\ \bibinfo {author} {\bibfnamefont {F.}~\bibnamefont {Nori}},\
  }\href {\doibase 10.1126/science.1177838} {\bibfield  {journal} {\bibinfo
  {journal} {Science}\ }\textbf {\bibinfo {volume} {326}},\ \bibinfo {pages}
  {108} (\bibinfo {year} {2009})}\BibitemShut {NoStop}%
\bibitem [{\citenamefont {Cirac}\ and\ \citenamefont
  {Zoller}(2012)}]{Cirac2012Goals}%
  \BibitemOpen
  \bibfield  {author} {\bibinfo {author} {\bibfnamefont {J.~I.}\ \bibnamefont
  {Cirac}}\ and\ \bibinfo {author} {\bibfnamefont {P.}~\bibnamefont {Zoller}},\
  }\href {http://dx.doi.org/10.1038/nphys2275} {\bibfield  {journal} {\bibinfo
  {journal} {Nat. Phys.}\ }\textbf {\bibinfo {volume} {8}},\ \bibinfo {pages}
  {264} (\bibinfo {year} {2012})}\BibitemShut {NoStop}%
\bibitem [{\citenamefont {Toda}(1991)}]{toda1991pp}%
  \BibitemOpen
  \bibfield  {author} {\bibinfo {author} {\bibfnamefont {S.}~\bibnamefont
  {Toda}},\ }\href {\doibase 10.1137/0220053} {\bibfield  {journal} {\bibinfo
  {journal} {SIAM J. Comput.}\ }\textbf {\bibinfo {volume} {20}},\ \bibinfo
  {pages} {865} (\bibinfo {year} {1991})}\BibitemShut {NoStop}%
\bibitem [{\citenamefont {Han}\ \emph {et~al.}(1997)\citenamefont {Han},
  \citenamefont {Hemaspaandra},\ and\ \citenamefont
  {Thierauf}}]{han1997threshold}%
  \BibitemOpen
  \bibfield  {author} {\bibinfo {author} {\bibfnamefont {Y.}~\bibnamefont
  {Han}}, \bibinfo {author} {\bibfnamefont {L.~A.}\ \bibnamefont
  {Hemaspaandra}}, \ and\ \bibinfo {author} {\bibfnamefont {T.}~\bibnamefont
  {Thierauf}},\ }\href {\doibase 10.1137/S0097539792240467} {\bibfield
  {journal} {\bibinfo  {journal} {SIAM J. Comput.}\ }\textbf {\bibinfo {volume}
  {26}},\ \bibinfo {pages} {59} (\bibinfo {year} {1997})}\BibitemShut {NoStop}%
\bibitem [{\citenamefont {Aaronson}(2005)}]{aaronson2005quantum}%
  \BibitemOpen
  \bibfield  {author} {\bibinfo {author} {\bibfnamefont {S.}~\bibnamefont
  {Aaronson}},\ }\href {\doibase 10.1098/rspa.2005.1546} {\bibfield  {journal}
  {\bibinfo  {journal} {Proc. R. Soc. A}\ }\textbf {\bibinfo {volume} {461}},\
  \bibinfo {pages} {3473} (\bibinfo {year} {2005})}\BibitemShut {NoStop}%
\bibitem [{\citenamefont {Hangleiter}\ \emph {et~al.}(2016)\citenamefont
  {Hangleiter}, \citenamefont {Kliesch}, \citenamefont {Schwarz},\ and\
  \citenamefont {Eisert}}]{hangleiter2016direct}%
  \BibitemOpen
  \bibfield  {author} {\bibinfo {author} {\bibfnamefont {D.}~\bibnamefont
  {Hangleiter}}, \bibinfo {author} {\bibfnamefont {M.}~\bibnamefont {Kliesch}},
  \bibinfo {author} {\bibfnamefont {M.}~\bibnamefont {Schwarz}}, \ and\
  \bibinfo {author} {\bibfnamefont {J.}~\bibnamefont {Eisert}},\ }\href@noop {}
  {\bibfield  {journal} {\bibinfo  {journal} {ArXiv e-prints}\ } (\bibinfo
  {year} {2016})},\ \Eprint {http://arxiv.org/abs/1602.00703} {arXiv:1602.00703
  [quant-ph]} \BibitemShut {NoStop}%
\bibitem [{\citenamefont {{Cramer}}\ \emph {et~al.}(2010)\citenamefont
  {{Cramer}}, \citenamefont {{Plenio}}, \citenamefont {{Flammia}},
  \citenamefont {{Somma}}, \citenamefont {{Gross}}, \citenamefont {{Bartlett}},
  \citenamefont {{Landon-Cardinal}}, \citenamefont {{Poulin}},\ and\
  \citenamefont {{Liu}}}]{cramer2010efficient}%
  \BibitemOpen
  \bibfield  {author} {\bibinfo {author} {\bibfnamefont {M.}~\bibnamefont
  {{Cramer}}}, \bibinfo {author} {\bibfnamefont {M.~B.}\ \bibnamefont
  {{Plenio}}}, \bibinfo {author} {\bibfnamefont {S.~T.}\ \bibnamefont
  {{Flammia}}}, \bibinfo {author} {\bibfnamefont {R.}~\bibnamefont {{Somma}}},
  \bibinfo {author} {\bibfnamefont {D.}~\bibnamefont {{Gross}}}, \bibinfo
  {author} {\bibfnamefont {S.~D.}\ \bibnamefont {{Bartlett}}}, \bibinfo
  {author} {\bibfnamefont {O.}~\bibnamefont {{Landon-Cardinal}}}, \bibinfo
  {author} {\bibfnamefont {D.}~\bibnamefont {{Poulin}}}, \ and\ \bibinfo
  {author} {\bibfnamefont {Y.-K.}\ \bibnamefont {{Liu}}},\ }\href {\doibase
  10.1038/ncomms1147} {\bibfield  {journal} {\bibinfo  {journal} {Nat.
  Commun.}\ }\textbf {\bibinfo {volume} {1}},\ \bibinfo {eid} {149} (\bibinfo
  {year} {2010})}\BibitemShut {NoStop}%
\bibitem [{\citenamefont {{Fefferman}}\ and\ \citenamefont
  {{Umans}}(2015)}]{fefferman2015power}%
  \BibitemOpen
  \bibfield  {author} {\bibinfo {author} {\bibfnamefont {B.}~\bibnamefont
  {{Fefferman}}}\ and\ \bibinfo {author} {\bibfnamefont {C.}~\bibnamefont
  {{Umans}}},\ }\href@noop {} {\bibfield  {journal} {\bibinfo  {journal} {ArXiv
  e-prints}\ } (\bibinfo {year} {2015})},\ \Eprint
  {http://arxiv.org/abs/1507.05592} {arXiv:1507.05592 [cs.CC]} \BibitemShut
  {NoStop}%
\bibitem [{\citenamefont {Aaronson}(2014)}]{aaronson2014equivalence}%
  \BibitemOpen
  \bibfield  {author} {\bibinfo {author} {\bibfnamefont {S.}~\bibnamefont
  {Aaronson}},\ }\href {\doibase 10.1007/s00224-013-9527-3} {\bibfield
  {journal} {\bibinfo  {journal} {Theor. Comp. Sys.}\ }\textbf {\bibinfo
  {volume} {55}},\ \bibinfo {pages} {281} (\bibinfo {year} {2014})}\BibitemShut
  {NoStop}%
\bibitem [{\citenamefont {Raussendorf}\ \emph {et~al.}(2006)\citenamefont
  {Raussendorf}, \citenamefont {Harrington},\ and\ \citenamefont
  {Goyal}}]{Raussendorf20062242}%
  \BibitemOpen
  \bibfield  {author} {\bibinfo {author} {\bibfnamefont {R.}~\bibnamefont
  {Raussendorf}}, \bibinfo {author} {\bibfnamefont {J.}~\bibnamefont
  {Harrington}}, \ and\ \bibinfo {author} {\bibfnamefont {K.}~\bibnamefont
  {Goyal}},\ }\href {\doibase http://dx.doi.org/10.1016/j.aop.2006.01.012}
  {\bibfield  {journal} {\bibinfo  {journal} {Ann. Phys.}\ }\textbf {\bibinfo
  {volume} {321}},\ \bibinfo {pages} {2242 } (\bibinfo {year}
  {2006})}\BibitemShut {NoStop}%
\bibitem [{\citenamefont {Briegel}\ \emph {et~al.}(2009)\citenamefont
  {Briegel}, \citenamefont {Browne}, \citenamefont {D{\"{u}}r}, \citenamefont
  {Raussendorf},\ and\ \citenamefont {{Van den
  Nest}}}]{Briegel2009Measurement}%
  \BibitemOpen
  \bibfield  {author} {\bibinfo {author} {\bibfnamefont {H.~J.}\ \bibnamefont
  {Briegel}}, \bibinfo {author} {\bibfnamefont {D.~E.}\ \bibnamefont {Browne}},
  \bibinfo {author} {\bibfnamefont {W.}~\bibnamefont {D{\"{u}}r}}, \bibinfo
  {author} {\bibfnamefont {R.}~\bibnamefont {Raussendorf}}, \ and\ \bibinfo
  {author} {\bibfnamefont {M.}~\bibnamefont {{Van den Nest}}},\ }\href
  {\doibase 10.1038/nphys1157} {\bibfield  {journal} {\bibinfo  {journal} {Nat.
  Phys.}\ }\textbf {\bibinfo {volume} {5}},\ \bibinfo {pages} {19} (\bibinfo
  {year} {2009})}\BibitemShut {NoStop}%
\bibitem [{\citenamefont {Raussendorf}\ and\ \citenamefont
  {Briegel}(2001)}]{Raussendorf2001A}%
  \BibitemOpen
  \bibfield  {author} {\bibinfo {author} {\bibfnamefont {R.}~\bibnamefont
  {Raussendorf}}\ and\ \bibinfo {author} {\bibfnamefont {H.~J.}\ \bibnamefont
  {Briegel}},\ }\href {\doibase 10.1103/PhysRevLett.86.5188} {\bibfield
  {journal} {\bibinfo  {journal} {Phys. Rev. Lett.}\ }\textbf {\bibinfo
  {volume} {86}},\ \bibinfo {pages} {5188} (\bibinfo {year}
  {2001})}\BibitemShut {NoStop}%
\bibitem [{\citenamefont {Raussendorf}\ \emph {et~al.}(2003)\citenamefont
  {Raussendorf}, \citenamefont {Browne},\ and\ \citenamefont
  {Briegel}}]{Raussendorf2003Measurement}%
  \BibitemOpen
  \bibfield  {author} {\bibinfo {author} {\bibfnamefont {R.}~\bibnamefont
  {Raussendorf}}, \bibinfo {author} {\bibfnamefont {D.~E.}\ \bibnamefont
  {Browne}}, \ and\ \bibinfo {author} {\bibfnamefont {H.~J.}\ \bibnamefont
  {Briegel}},\ }\href {\doibase 10.1103/PhysRevA.68.022312} {\bibfield
  {journal} {\bibinfo  {journal} {Phys. Rev. A}\ }\textbf {\bibinfo {volume}
  {68}},\ \bibinfo {pages} {022312} (\bibinfo {year} {2003})}\BibitemShut
  {NoStop}%
\bibitem [{\citenamefont {Broadbent}\ \emph {et~al.}(2009)\citenamefont
  {Broadbent}, \citenamefont {Fitzsimons},\ and\ \citenamefont
  {Kashefi}}]{broadbent2009universal}%
  \BibitemOpen
  \bibfield  {author} {\bibinfo {author} {\bibfnamefont {A.}~\bibnamefont
  {Broadbent}}, \bibinfo {author} {\bibfnamefont {J.}~\bibnamefont
  {Fitzsimons}}, \ and\ \bibinfo {author} {\bibfnamefont {E.}~\bibnamefont
  {Kashefi}},\ }in\ \href {\doibase 10.1109/FOCS.2009.36} {\emph {\bibinfo
  {booktitle} {Foundations of Computer Science, 2009. FOCS '09. 50th Annual
  IEEE Symposium on}}}\ (\bibinfo {year} {2009})\ pp.\ \bibinfo {pages}
  {517--526}\BibitemShut {NoStop}%
\bibitem [{\citenamefont {Kitaev}(1997)}]{kitaev1997quantum}%
  \BibitemOpen
  \bibfield  {author} {\bibinfo {author} {\bibfnamefont {A.~Y.}\ \bibnamefont
  {Kitaev}},\ }\href {http://stacks.iop.org/0036-0279/52/i=6/a=R02} {\bibfield
  {journal} {\bibinfo  {journal} {Russ. Math. Surv.}\ }\textbf {\bibinfo
  {volume} {52}},\ \bibinfo {pages} {1191} (\bibinfo {year}
  {1997})}\BibitemShut {NoStop}%
\bibitem [{\citenamefont {Jaksch}\ \emph {et~al.}(1999)\citenamefont {Jaksch},
  \citenamefont {Briegel}, \citenamefont {Cirac}, \citenamefont {Gardiner},\
  and\ \citenamefont {Zoller}}]{Jaksch1999Entanglement}%
  \BibitemOpen
  \bibfield  {author} {\bibinfo {author} {\bibfnamefont {D.}~\bibnamefont
  {Jaksch}}, \bibinfo {author} {\bibfnamefont {H.-J.}\ \bibnamefont {Briegel}},
  \bibinfo {author} {\bibfnamefont {J.~I.}\ \bibnamefont {Cirac}}, \bibinfo
  {author} {\bibfnamefont {C.~W.}\ \bibnamefont {Gardiner}}, \ and\ \bibinfo
  {author} {\bibfnamefont {P.}~\bibnamefont {Zoller}},\ }\href {\doibase
  10.1103/PhysRevLett.82.1975} {\bibfield  {journal} {\bibinfo  {journal}
  {Phys. Rev. Lett.}\ }\textbf {\bibinfo {volume} {82}},\ \bibinfo {pages}
  {1975} (\bibinfo {year} {1999})}\BibitemShut {NoStop}%
\bibitem [{\citenamefont {Mandel}\ \emph {et~al.}(2003)\citenamefont {Mandel},
  \citenamefont {Greiner}, \citenamefont {Widera}, \citenamefont {Rom},
  \citenamefont {Hansch},\ and\ \citenamefont {Bloch}}]{Mandel2003Controlled}%
  \BibitemOpen
  \bibfield  {author} {\bibinfo {author} {\bibfnamefont {O.}~\bibnamefont
  {Mandel}}, \bibinfo {author} {\bibfnamefont {M.}~\bibnamefont {Greiner}},
  \bibinfo {author} {\bibfnamefont {A.}~\bibnamefont {Widera}}, \bibinfo
  {author} {\bibfnamefont {T.}~\bibnamefont {Rom}}, \bibinfo {author}
  {\bibfnamefont {T.~W.}\ \bibnamefont {Hansch}}, \ and\ \bibinfo {author}
  {\bibfnamefont {I.}~\bibnamefont {Bloch}},\ }\href
  {http://dx.doi.org/10.1038/nature02008} {\bibfield  {journal} {\bibinfo
  {journal} {Nature}\ }\textbf {\bibinfo {volume} {425}},\ \bibinfo {pages}
  {937} (\bibinfo {year} {2003})}\BibitemShut {NoStop}%
\bibitem [{\citenamefont {Weitenberg}\ \emph {et~al.}(2011)\citenamefont
  {Weitenberg}, \citenamefont {Endres}, \citenamefont {Sherson}, \citenamefont
  {Cheneau}, \citenamefont {Schausz}, \citenamefont {Fukuhara}, \citenamefont
  {Bloch},\ and\ \citenamefont {Kuhr}}]{Weitenberg2011Nature}%
  \BibitemOpen
  \bibfield  {author} {\bibinfo {author} {\bibfnamefont {C.}~\bibnamefont
  {Weitenberg}}, \bibinfo {author} {\bibfnamefont {M.}~\bibnamefont {Endres}},
  \bibinfo {author} {\bibfnamefont {J.~F.}\ \bibnamefont {Sherson}}, \bibinfo
  {author} {\bibfnamefont {M.}~\bibnamefont {Cheneau}}, \bibinfo {author}
  {\bibfnamefont {P.}~\bibnamefont {Schausz}}, \bibinfo {author} {\bibfnamefont
  {T.}~\bibnamefont {Fukuhara}}, \bibinfo {author} {\bibfnamefont
  {I.}~\bibnamefont {Bloch}}, \ and\ \bibinfo {author} {\bibfnamefont
  {S.}~\bibnamefont {Kuhr}},\ }\href {http://dx.doi.org/10.1038/nature09827}
  {\bibfield  {journal} {\bibinfo  {journal} {Nature}\ }\textbf {\bibinfo
  {volume} {471}},\ \bibinfo {pages} {319} (\bibinfo {year}
  {2011})}\BibitemShut {NoStop}%
\bibitem [{\citenamefont {{Zupancic}}\ \emph {et~al.}(2016)\citenamefont
  {{Zupancic}}, \citenamefont {{Preiss}}, \citenamefont {{Ma}}, \citenamefont
  {{Lukin}}, \citenamefont {{Tai}}, \citenamefont {{Rispoli}}, \citenamefont
  {{Islam}},\ and\ \citenamefont {{Greiner}}}]{Zupancic2016Ultra}%
  \BibitemOpen
  \bibfield  {author} {\bibinfo {author} {\bibfnamefont {P.}~\bibnamefont
  {{Zupancic}}}, \bibinfo {author} {\bibfnamefont {P.~M.}\ \bibnamefont
  {{Preiss}}}, \bibinfo {author} {\bibfnamefont {R.}~\bibnamefont {{Ma}}},
  \bibinfo {author} {\bibfnamefont {A.}~\bibnamefont {{Lukin}}}, \bibinfo
  {author} {\bibfnamefont {M.~E.}\ \bibnamefont {{Tai}}}, \bibinfo {author}
  {\bibfnamefont {M.}~\bibnamefont {{Rispoli}}}, \bibinfo {author}
  {\bibfnamefont {R.}~\bibnamefont {{Islam}}}, \ and\ \bibinfo {author}
  {\bibfnamefont {M.}~\bibnamefont {{Greiner}}},\ }\href@noop {} {\bibfield
  {journal} {\bibinfo  {journal} {ArXiv e-prints}\ } (\bibinfo {year}
  {2016})},\ \Eprint {http://arxiv.org/abs/1604.07653} {arXiv:1604.07653
  [cond-mat.quant-gas]} \BibitemShut {NoStop}%
\bibitem [{\citenamefont {Bakr}\ \emph {et~al.}(2010)\citenamefont {Bakr},
  \citenamefont {Peng}, \citenamefont {Tai}, \citenamefont {Ma}, \citenamefont
  {Simon}, \citenamefont {Gillen}, \citenamefont {F{\"o}lling}, \citenamefont
  {Pollet},\ and\ \citenamefont {Greiner}}]{Bakr2010Probing}%
  \BibitemOpen
  \bibfield  {author} {\bibinfo {author} {\bibfnamefont {W.~S.}\ \bibnamefont
  {Bakr}}, \bibinfo {author} {\bibfnamefont {A.}~\bibnamefont {Peng}}, \bibinfo
  {author} {\bibfnamefont {M.~E.}\ \bibnamefont {Tai}}, \bibinfo {author}
  {\bibfnamefont {R.}~\bibnamefont {Ma}}, \bibinfo {author} {\bibfnamefont
  {J.}~\bibnamefont {Simon}}, \bibinfo {author} {\bibfnamefont {J.~I.}\
  \bibnamefont {Gillen}}, \bibinfo {author} {\bibfnamefont {S.}~\bibnamefont
  {F{\"o}lling}}, \bibinfo {author} {\bibfnamefont {L.}~\bibnamefont {Pollet}},
  \ and\ \bibinfo {author} {\bibfnamefont {M.}~\bibnamefont {Greiner}},\ }\href
  {\doibase 10.1126/science.1192368} {\bibfield  {journal} {\bibinfo  {journal}
  {Science}\ }\textbf {\bibinfo {volume} {329}},\ \bibinfo {pages} {547}
  (\bibinfo {year} {2010})}\BibitemShut {NoStop}%
\bibitem [{\citenamefont {Sherson}\ \emph {et~al.}(2010)\citenamefont
  {Sherson}, \citenamefont {Weitenberg}, \citenamefont {Endres}, \citenamefont
  {Cheneau}, \citenamefont {Bloch},\ and\ \citenamefont
  {Kuhr}}]{Sherson2010Single}%
  \BibitemOpen
  \bibfield  {author} {\bibinfo {author} {\bibfnamefont {J.~F.}\ \bibnamefont
  {Sherson}}, \bibinfo {author} {\bibfnamefont {C.}~\bibnamefont {Weitenberg}},
  \bibinfo {author} {\bibfnamefont {M.}~\bibnamefont {Endres}}, \bibinfo
  {author} {\bibfnamefont {M.}~\bibnamefont {Cheneau}}, \bibinfo {author}
  {\bibfnamefont {I.}~\bibnamefont {Bloch}}, \ and\ \bibinfo {author}
  {\bibfnamefont {S.}~\bibnamefont {Kuhr}},\ }\href
  {http://dx.doi.org/10.1038/nature09378} {\bibfield  {journal} {\bibinfo
  {journal} {Nature}\ }\textbf {\bibinfo {volume} {467}},\ \bibinfo {pages}
  {68} (\bibinfo {year} {2010})}\BibitemShut {NoStop}%
\bibitem [{\citenamefont {Kitaev}\ \emph {et~al.}(2002)\citenamefont {Kitaev},
  \citenamefont {Shen},\ and\ \citenamefont {Vyalyi}}]{kitaev2002classical}%
  \BibitemOpen
  \bibfield  {author} {\bibinfo {author} {\bibfnamefont {A.~Y.}\ \bibnamefont
  {Kitaev}}, \bibinfo {author} {\bibfnamefont {A.}~\bibnamefont {Shen}}, \ and\
  \bibinfo {author} {\bibfnamefont {M.~N.}\ \bibnamefont {Vyalyi}},\
  }\href@noop {} {\emph {\bibinfo {title} {Classical and quantum
  computation}}},\ Vol.~\bibinfo {volume} {47}\ (\bibinfo  {publisher}
  {American Mathematical Society Providence},\ \bibinfo {year}
  {2002})\BibitemShut {NoStop}%
\bibitem [{\citenamefont {{Preskill}}(2012)}]{preskill2012quantum}%
  \BibitemOpen
  \bibfield  {author} {\bibinfo {author} {\bibfnamefont {J.}~\bibnamefont
  {{Preskill}}},\ }\href@noop {} {\bibfield  {journal} {\bibinfo  {journal}
  {ArXiv e-prints}\ } (\bibinfo {year} {2012})},\ \Eprint
  {http://arxiv.org/abs/1203.5813} {arXiv:1203.5813 [quant-ph]} \BibitemShut
  {NoStop}%
\end{thebibliography}
\end{document}